\documentclass{article}

\usepackage{arxiv}

\usepackage[utf8]{inputenc} 
\usepackage[T1]{fontenc}    
\usepackage{url}            
\usepackage{booktabs}       
\usepackage{amsfonts}       
\usepackage{nicefrac}
\usepackage{graphicx}
\usepackage[numbers,sort&compress]{natbib}
\usepackage{amsthm,amsmath,amssymb}
\usepackage{mathrsfs}
\usepackage{doi}

\title{Topological states in quasicrystals}

\author{ \href{https://orcid.org/0000-0003-1311-1730}{\includegraphics[scale=0.06]{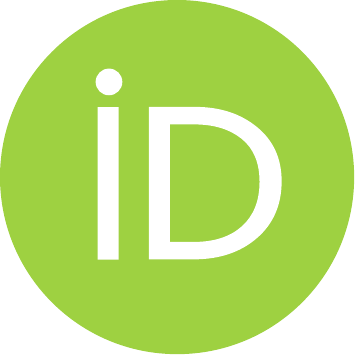}\hspace{1mm}Jiahao Fan} \\
	School of Physics,Peking University\\
	Center for High Energy Physics\\
	Collaborative Innovation Center of Quantum Matter\\
	Beijing, China 100871 \\
	\texttt{jiahaofan@pku.edu.cn} \\
	\And
	\href{https://orcid.org/0000-0002-0283-8603}{\includegraphics[scale=0.06]{orcid.pdf}\hspace{1mm}Huaqing Huang} \\
	School of Physics,Peking University\\
	Center for High Energy Physics\\
	Collaborative Innovation Center of Quantum Matter\\
	Beijing, China 100871 \\
	\texttt{huaqing.huang@pku.edu.cn} \\
}

\renewcommand{\shorttitle}{Topological states in quasicrystals}

\hypersetup{
pdftitle={Topological states in quasicrystals},
pdfsubject={q-bio.NC, q-bio.QM},
pdfauthor={David S.~Hippocampus, Elias D.~Striatum},
pdfkeywords={First keyword, Second keyword, More},
}

\begin{document}
\maketitle
\tableofcontents

\newpage
\begin{abstract}
With the rapid development of topological states in crystals, the study of topological states has been extended to quasicrystals in recent years. In this review, we summarize the recent progress of topological states in quasicrystals, particularly focusing on one-dimensional (1D) and 2D systems. We first give a brief introduction to quasicrystalline structures. Then, we discuss topological phases in 1D quasicrystals where the topological nature is attributed to the synthetic dimensions associated with the quasiperiodic order of quasicrystals. We further present the generalization of various types of crystalline topological states to 2D quasicrystals, where real-space expressions of corresponding topological invariants are introduced due to the lack of translational symmetry in quasicrystals. Finally, since quasicrystals possess forbidden symmetries in crystals such as five-fold and eight-fold rotation, we provide an overview of unique quasicrystalline symmetry-protected topological states without crystalline counterpart.
\end{abstract}


\section{Introduction}
\noindent Topological states of matter are among the most intriguing research topics of condensed-matter physics during the last half century \cite{qi2011topological,RevModPhys.82.3045,bernevig2013topological}. Historically, the integer quantum Hall effect (IQHE) discovered in the 1980s \cite{Klitzing} is an early example of topological states where the quantized Hall conductance was later explained by the Thouless-Kohmoto-Nightingale-den Nijs (TKNN) integer of occupied energy bands (a topological invariant also known as the Chern number) \cite{Thouless}. Recently, the discovery of topological insulators and superconductors and their classification for the ten Altland-Zirnbauer symmetry classes based on internal discrete symmetries (time-reversal, particle-hole, and chiral symmetry) \cite{altland1997nonstandard,tenfold_way1,kitaev2009periodic,tenfold_way2} have stimulated various theoretical and experimental studies of real topological materials \cite{zhang2013topological,ando2013topological,sato2017topological,huanghqCMS,huanghqIIIV,huanghqSemiDirac,huang2020unified}. The subsequent development of topological classification has been extended to materials in which the crystal structure is essential for the protection of topological phases. This includes topological crystalline insulators (TCIs) \cite{fu2011cry,ando2015topological}, in which the topological nature of electronic structures arises from crystal symmetries (such as the mirror or rotational symmetries), and higher-order topological insulators (HOTIs) \cite{schindler2018higher,langbehn2017reflection}, in which topologically protected gapless states only occur at the intersection of crystal facets, but is gapped otherwise. More recently, the theory of symmetry-based indicator has been established to diagnose the underlying band topology of a large number of crystalline materials via a high-throughput computational search in crystalline material databases \cite{po2017symmetry,bradlyn2017topological,song2018quantitative,zhang2019catalogue,tang2019comprehensive, vergniory2019complete}.

The rapid progress of topological states in crystals has sparked considerable interest in the study of topological states in aperiodic systems, such as quasicrystals \cite{PhysRevLett.53.1951, steinhardt1987physics, janot1994quasicrystals}.
Recent advances have shown various analogous topological states in quasicrystals as well as novel topological quasicrystalline states without crystalline counterparts \cite{Kraus,Kraus2,Kraus3,Verbin,Tran,Hofstadter,Hofstadter2,Hofstadter3,PhysRevB.101.115413,Huang,Huang1,Huang3,TAI,TAI1,TAI2,He,Huang2,HOTI1,Chen,HOTI2,HOTI3,floquet1,floquet2,tezuka2012reentrant,degottardi2013majorana,ghadimi2017majorana,Loring,TSC1,TSC2,li2020quantum,yao2018quasicrystalline}. Although the unique structural feature of quasicrystals leads to significant difficulties in dealing with topological states, it also opens up new opportunities to realize exotic topological phenomena that are impossible in crystals.

Quasicrystals, which are special kinds of matter that possess a long-range orientational order but no translational symmetry, have attached widespread interest since their first discovery in 1982 \cite{PhysRevLett.53.1951}. Due to the lack of translational symmetry, the Bloch theorem and, therefore, the topological band theory \cite{RevModPhys.88.021004} does not apply to quasicrystals, which prohibits an intuitive analysis of the prevailing band-inversion mechanism for topological states in quasicrystals. Moreover, most topological invariants, such as the $\mathbb{Z}_2$ index, are defined only for periodic systems \cite{kane2005z,fu2007inv}. Thus, in order to identify topological states in quasicrystals, new topological invariants that apply to aperiodic systems are in urgent need.

On the other hand, the long-range quasicrystalline order can be seen as originating from periodic structures of a dimension higher than the physical one. For example, the one-dimensional (1D) Fibonacci quasicrystals can be described as a projection of a 2D lattice on a line with an irrational slope (see Fig.~\ref{projection}) \cite{janot1994quasicrystals,jagannathan2020fibonacci}. Remnants of the higher dimensionality appear as additional degrees of freedom in the form of shifts of the origin of the quasiperiodic order, and can lead to virtual topological properties in dimensions higher than the spatial one, with some ``synthetic dimensions'' occurring in parameter spaces \cite{prodan2015virtual}. For example, it was recently shown that 1D quasicrystals are assigned Chern numbers and exhibit topological properties of the 2D IQHE \cite{Kraus,Kraus2}. More interestingly, beyond the classical crystallographic restriction of periodicity, quasicrystals contain unique symmetries (such as five-fold and eight-fold rotations) that are forbidden in conventional crystals \cite{PhysRevLett.53.2477,PhysRevLett.59.1010, wang1988symmetry}. These novel symmetries can lead to new types of topological states that have no crystalline counterparts \cite{HOTI1,Chen, HOTI2,HOTI3}.

Here we review these recent developments of topological states in quasicrystals. We first give a concise introduction to quasicrystalline structures in Sec. 2, discussing the fundamental definition and the construction of quasicrystals. Then we summarize the topological aspects of quasicrystals in different dimensions. In Sec. 3, we focus on 1D quasicrystals where topological properties are attributed to the synthetic dimensions associated with the quasiperiodic order of quasicrystals. In Sec. 4, we discuss the generalization of topological states existing in crystals to quasicrystals where different real-space topological invariants are proposed to characterize these quasicrystalline topological counterparts. In Sec. 5, we introduce the unique topological states that are protected by quasicrystalline rotational symmetries. We end with a short summary of the current status of this field and outline some promising directions for future research in Sec. 6.

\section{Structures of quasicrystals}
\noindent Quasicrystal was a special form of solid matter that is ordered but not periodic. The quasicrystalline structure was first observed in rapidly solidified aluminium-manganese alloys (Al$_6$Mn) via X-ray diffraction experiments \cite{PhysRevLett.53.1951}. The key experimental feature was the ten-fold diffraction patterns with sharp Bragg peaks, indicating a long-range order that possesses a crystallographically disallowed rotational symmetry which precluded periodicity. The discovery enables the redefinition of ``order" in crystalline materials, which is no longer just ``periodic" as in the translationally invariant crystals. Rather it is suggested that one should use an operational definition for an ordered structure: a translationally ordered structure is a structure whose scattering amplitude is given by a discrete sum of Bragg peaks. \cite{steinhardt1987physics}

With this definition which includes both crystals and quasicrystals, one further defines that a crystal in $d$ dimensions is a translationally ordered structure with a basis (to index the Bragg peaks) whose rank is equal to $d$, while a quasiperiodic structure in $d$ dimensions is a translationally ordered structure with a finite basis whose rank exceeds $d$ \cite{steinhardt1987physics}. Steinhardt further proposed a definition to classify quasiperiodic structures into incommensurate crystals and quasicrystals according to whether it has crystallographically disallowed orientational symmetry, such as the five-fold, eight-fold, ten-fold, and 12-fold rotational symmetry \cite{steinhardt1987physics}. However, given the similarity between the two, in this review, we use Lifshiz's definition which drops the forbidden symmetry condition \cite{lifshitz2003quasicrystals}, and treats quasicrystals just as an abbreviation of quasiperiodic structures.

\begin{figure}[hbt]
\centering
\includegraphics[width=0.5\columnwidth]{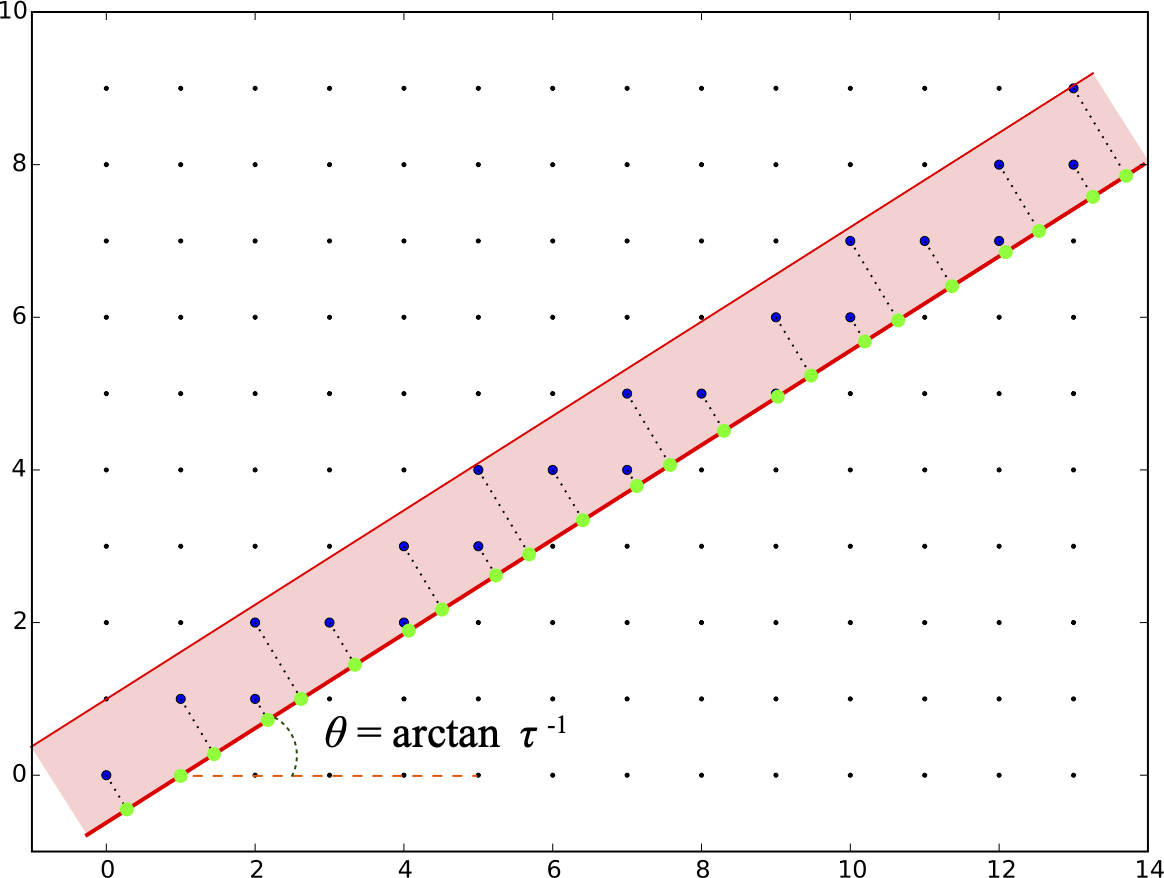}
\caption{Illustration of the projection method for the construction of the Fibonacci lattice. The shaded area represents the projection window, the lattice sites in the projection window are all projected perpendicularly onto the red line with a slope of $\tau^{-1}=2/(\sqrt{5}+1)$, the resulting green dots form the quasiperiodic lattice sites. }
\label{projection}
\end{figure}

In the study of electronic properties of a quasicrystal, there are generally two ways of building a model \cite{steinhardt1987physics}: one is the tight-binding approach, which explicitly constructs a quasicrystalline lattice and builds a tight-binding model based on such a lattice. The other is the ``density wave" approach \cite{Bak1985}, which considers the ionic lattice as an incommensurate sum of plane waves according to the diffraction pattern. In this review, we mainly adopt the first approach to build models for quasicrystals. To construct a quasicrystalline lattice, one can use the so-called ``projection'' method \cite{PhysRevLett.54.2688,PhysRevLett.55.2883}, in which quasicrystals are generated by the projection from a higher-dimensional periodic lattice with the projection hyperplane at an incommensurate orientation. As shown in Fig.~\ref{projection}, the relation of quasicrystals to higher-dimensional crystals is illustrated by the construction of the 1D Fibonacci quasicrystal, which is obtained by applying the projection procedure on a square lattice onto the line with a slope of $\tau^{-1}=2/(1+\sqrt{5})$. This relation between quasicrystals and higher-dimensional crystals can be generalized to 2D and 3D quasicrystals.

The projection process manifests the fact that the rank of a quasicrystal in $d$ dimensions exceeds $d$. Furthermore, the quasiperiodic structure has unique low-energy excitations due to the rearrangement of atoms, called ``phasons'' \cite{PhysRevB.34.3345}, which are described as phonons in superspace. Similar to phonons, phasons are associated with the change of atomic position. While phonons are related to atomic motion in physical space, phasons actually deforms the slice of projection that creates the quasiperiodic structure, therefore, are excitations in superspace. Remnants of the higher dimensionality provide additional phasonic degrees of freedom which can induce measurable topological effects on the lower-dimensional quasiperiodic systems. This is also the essential idea for the realization of topological states in quasicrystals using synthetic dimensions, which will discuss in Sec. 3.

Since the experimental study of real quasicrystal materials is largely obstructed by the rigorous growth processes, stability conditions, and the unavoidable presence of disorder \cite{poon1992electronic}, there are also alternative experimental efforts to study the properties of quasicrystals based on artificial quasicrystal structures, such as quasicrystalline optical lattices \cite{PhysRevLett.79.3363,guidoni1999atomic,corcovilos2019two,PhysRevLett.122.110404,PhysRevLett.125.200604}, photonic and phononic quasicrystalline metacrystals \cite{steurer2007photonic,kaliteevski2000two,freedman2006wave}. Comparing to real quasicrystals, these composite systems with a quasicrystalline order can be easily designed and fabricated, which provides an ideal platform for investigating quasicrystals. Similar to studying the propagation of electrons in quasicrystals, the propagation of ultracold atoms, light, or sound through composite systems with quasicrystalline order are investigated in these artificial quasicrystals \cite{guidoni1999atomic,jagannathan2014eightfold,Kraus, verbin2015topological, bayindir2001photonic,della2005band,PhysRevX.7.041047,PhysRevLett.119.260401,PhysRevLett.80.956,PhysRevLett.90.055501,zoorob2000complete}.

In the quasicrystalline optical lattices for ultracold atoms, the optical lattice is formed by the laser interference pattern which traps ultracold atoms. The optical potential is given by the time-averaged electrical field intensity\cite{bloch2005ultracold}
\begin{equation}
V_{\mathrm{opt}}(\boldsymbol{r})=\alpha\left\langle|\boldsymbol{E}(\boldsymbol{r}, t)|^{2}\right\rangle_{t},
\end{equation}
where $\alpha=3 \pi c^{3} \epsilon_{0} \Gamma / \omega_{0}^{3} \delta,$ with $\Gamma$ being the spontaneous scattering rate, and $\delta \equiv \omega-\omega_{0}$ is the detuning of the laser frequency $\omega$ from the atomic resonance $\omega_{0}$. By adjusting the laser interference pattern, one can simulate different types of quasicrystals and study novel behaviors of atoms in the ``synthetic'' quasicrystalline lattices. For example, the dynamics of ultracold atoms in five-fold \cite{guidoni1999atomic,corcovilos2019two} and eight-fold symmetric quasicrystalline optical lattices \cite{jagannathan2014eightfold,PhysRevLett.122.110404,PhysRevLett.125.200604} have been studied experimentally.

The photonic metacrystals are also widely used to implement different quasicrystal models, such as the Harper \cite{Kraus}, Fibonacci \cite{verbin2015topological}, and Penrose model \cite{bayindir2001photonic,della2005band}. Because the time evolution of the Schr\"{o}dinger equation for electrons can be approximately mapped to the paraxial wave equation for the propagation of light through a photonic metacrystal with varying refractive index, which is given by \cite{ozawa2019topological}
\begin{equation}
\mathrm{i} \partial_{z} E=-\frac{1}{2 k} \nabla_{\perp}^{2} E-\frac{k \Delta n}{n_{0}} E,
\label{Hphotonic}
\end{equation}
where $\nabla_{\perp} \equiv$ $\partial_{x}^{2}+\partial_{y}^{2}$, $k$ is the wavevector of the electric field, and $E$ is the envelope of the electric field $\boldsymbol{E}(x, y, z)=\hat{z} E(x, y, z) \exp (\mathrm{i} k z)$ within the paraxial approximation\cite{rechtsman2013photonic}. $\Delta n$ and $n_{0}$ are the varying and average components of the refractive index $n$, respectively. Therefore, the propagation of photons in photonic quasicrystals, which is affected by the quasicrystalline nanostructure with spatially varying refractive index, can simulate the motion of electrons in quasicrystals. Since it is relatively easy to fabricate photonic quasicrystals, of which a variety of parameters are controllable, photonic quasicrystals offer a powerful synthetic platform to study quasicrystals.

\section{Realization of topological state in quasicrystals using synthetic dimensions}
\noindent The surprising connection between quasicrystals and topological states of matter was initially discovered in 1D quasicrystals by Kraus and colleagues \cite{Kraus}. As mentioned above, 1D quasicrystals can be described by a projection of 2D crystals. If the Hamiltonian of a 1D quasicrystal system depends on a periodic parameter, then this parameter can be considered as a synthetic dimension. The effective 2D ancestor system may have a nontrivial topological invariant (such as the Chern number in 2D IQHE) and, therefore, give rise to topological phenomena in the 1D descendant quasicrystal. A comprehensive review of topology in 1D quasicrystals is already available \cite{zilberberg2020topology}. Here we briefly introduce the basic idea for the realization of topological states in quasicrystal using synthetic dimensions.

\subsection{Topological states in 1D quasiperiodic structures\label{IIIA}}
\noindent In 2012, Kraus \textit{et al.} established the mathematical connection between 1D quasicrystals and 2D IQHE by considering a 1D quasicrystal model given by the general tight-binding Hamiltonian with nearest-neighbor hopping and an on-site potential \cite{Kraus,Kraus2}
\begin{equation}
H=\sum_{n}\left[\left(t+\lambda^{\mathrm{od}} V_{n}^{\mathrm{od}}\right) c_{n}^{\dagger} c_{n+1}+\mathrm{H.c.}+\lambda^{\mathrm{d}} V_{n}^{\mathrm{d}} c_{n}^{\dagger} c_{n}\right],
\label{eq:generalized 1D}
\end{equation}
where $c_n$ is the single-particle annihilation operator at site $n$, $t$ is the hopping amplitude, the real and positive parameters $\lambda^{\mathrm{od}}$ and $\lambda^\mathrm{d}$ control the strength of off-diagonal and diagonal potential modulation, respectively. The quasiperiodicity of different quasicrystal models is encoded in potential modulations $V_n^{\mathrm{od}}$ and $V_n^\mathrm{d}$.

\begin{figure}[hbt]
\centering
\includegraphics[width=6cm]{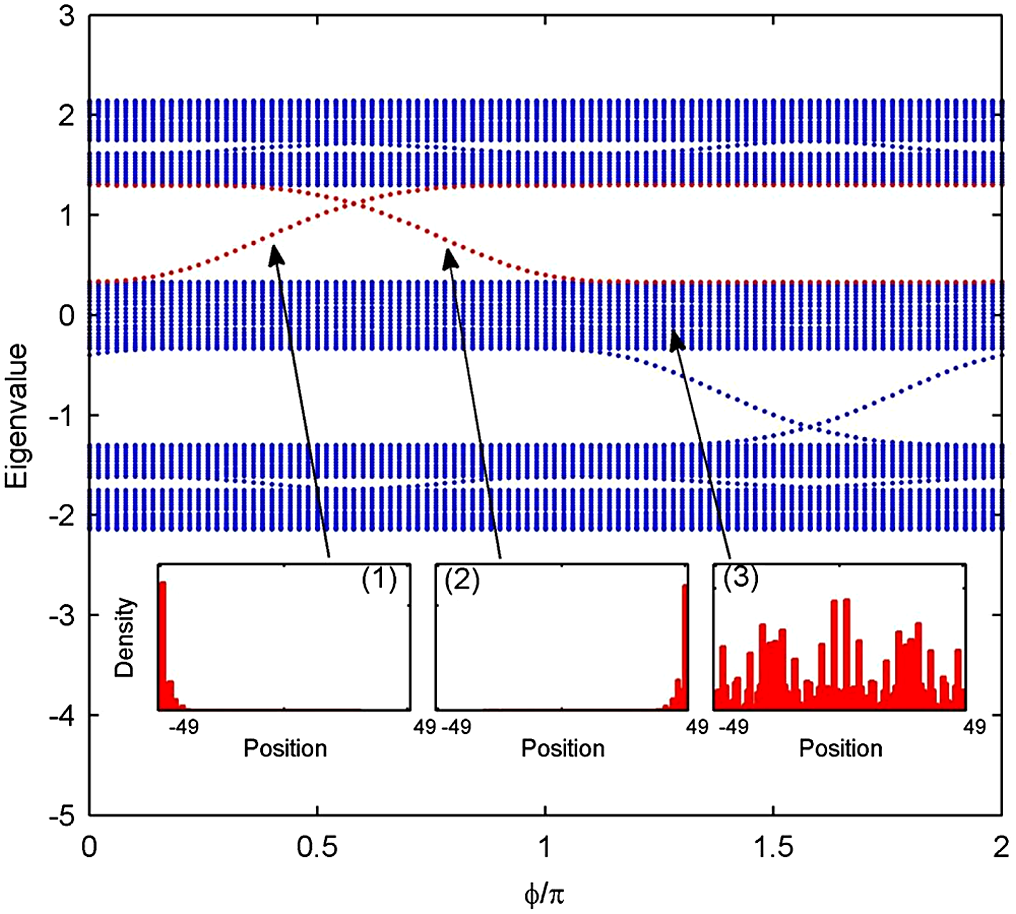}
\caption{The numerically calculated spectrum of diagonal Harper as a function of the phase $\phi$ for $t=1, \lambda=0.5$, $b=(\sqrt{5}+1) / 2$ (the golden mean), and $n=-49 \ldots 49$. The insets show the boundary states and bulk states of the model. Reproduced from Ref.~\cite{Kraus}}
\label{spectrum}
\end{figure}
They first considered the Harper model (also known as the Aubry-Andr\'{e} model) \cite{aubry1980analyticity} whose potential modulation can be written as $V_{n}^{\mathrm{H}}(\phi)=\cos (2 \pi b n+\phi)$, where $b$ controls the periodicity of the modulation and $\phi$ is the modulation phase. Specifically, by setting $\lambda^\mathrm{od}=0, \lambda^\mathrm{d}\neq0$ and $V_n^\mathrm{d} = V_{n}^{\mathrm{H}}(\phi)$  in Eq.~(\ref{eq:generalized 1D}), one obtains the diagonal Harper model \cite{Kraus},
\begin{equation}
H (\phi)=\sum_{n} t  c_{n}^{\dagger} c_{n+1}+\mathrm{H.c.}+\lambda^{\mathrm{d}} \cos (2 \pi b n+\phi) c_{n}^{\dagger} c_{n}.
\label{harper}
\end{equation}
Whenever $b$ is irrational, the modulation is incommensurate with the lattice and describes a quasicrystal. 
The modulation phase $\phi$, which encodes a ``phasonic'' continuous shit of the potential, serves as an additional dimension, and it plays a crucial role in constructing the mapping between 1D quasicrystals and 2D IQHEs.
The energy spectrum of the diagonal Harper model as a function of $\phi$ is presented in Fig.~\ref{spectrum}. Although the bulk bands are almost unchanged, boundary states, which are the physical manifestation of the nontrivial topological phase, are found in the bulk gap.

\begin{figure}[hbt]
\centering
\includegraphics[width=0.5\columnwidth]{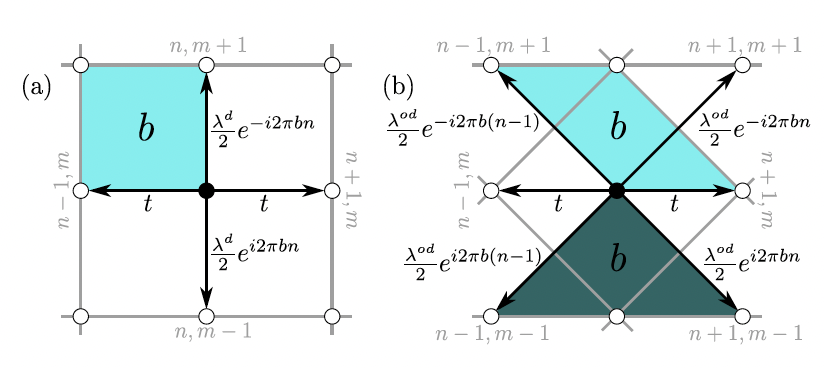}
\caption{2D lattice ancestor Hamiltonians. The electrons hop on a rectangular lattice in the presence of a perpendicular magnetic field with $b$ flux quantum per unit cell. (a) 2D ancestor Hamiltonian of diagonal Harper model. (b) 2D ancestor Hamiltonian of off-diagonal Harper model. Reproduced from Ref.~\cite{Kraus2}}
\label{2D IQHE}
\end{figure}

To map the diagonal Harper model to the lattice version of the 2D IQHE, they consider $\phi$ as the dimensionless crystal momentum $ka$, where $a$ is the lattice spacing along the second dimension. Therefore, for any given $\phi$, Eq.~(\ref{harper}) can be viewed as the $k$-th Fourier component of a 2D ancestor Hamiltonian. By performing an inverse Fourier transformation with respect to $\phi$, one can obtain the 2D real-space ancestor model \cite{Kraus2},
\begin{equation}
\mathcal{H}_{\mathrm{d}} =\sum_{n, m}\left[t c_{n, m}^{\dagger} c_{n+1, m}+\frac{\lambda^{\mathrm{d}}}{2} e^{i 2 \pi b n} c_{n, m}^{\dagger} c_{n, m+1}+\mathrm{H.c.}\right],
\end{equation}
where $m$ represents lattice site of the second dimension. This model describes electrons hopping on a 2D rectangular lattice in the presence of a uniform perpendicular magnetic field with $b$ flux quantum per unit cell, 
as shown in Fig.~\ref{2D IQHE}(a).

Due to the presence of magnetic field which breaks the lattice translational symmetry, the magnetic translation group can be introduced \cite{zak1964magnetic}. It is generated by the operators $T_{\hat{m}}$ and $T_{\hat{n}}$ where $T_{\hat{m}} c_{n,m} T_{\hat{m}}^{-1}=c_{n,m+1}$ and $T_{\hat{n}} c_{n,m} T_{\hat{n}}^{-1} =e^{-i2\pi b m}c_{n+1,m}$. For a rational flux $b=p/q$, it is possible to diagonalize simultaneously $\mathcal{H}_{\mathrm{d}}, T_{q\hat{n}}=(T_{\hat{n}})^q$ and $T_{\hat{m}}$, as they commute with each other. It actually corresponds to an enlarged real-space magnetic unit cell, which gives rise to the spectrum composed of $q$ bands \cite{Hofstadter}. As demonstrated by Thouless \textit{et al.}\cite{Thouless}, each gap in the spectrum  is associated with a quantized Chern number $\nu$ (Hall conductance $\sigma_H=\nu e^2/h$).
Specifically, the Chern number associated with the gap $r=1,\cdots, (q-1)$ in the spectrum is given by
\begin{equation}
\nu_r=\frac{1}{2\pi i}\int^{2\pi}_0 d\phi d\theta C_r(\phi,\theta),
\label{chern}
\end{equation}
where $\phi$ and $\theta$ are phase twists associated with the periodic boundary conditions (PBCs) for the 2D ancestor model, and
\begin{equation}
C_r(\phi,\theta)=\mathrm{Tr}\left(P_r\left[\frac{\partial P_r}{\partial \phi}, \frac{\partial P_r}{\partial \theta}\right]\right),
\end{equation}
is the Chern density with $P_r(\phi,\theta)=\sum_{E_n<E_{\mathrm{gap}}^r}|n\rangle\langle n|$ being the projector on the state below the gap $r$. Alternatively, due to the symmetry of the magnetic translation group, the Chern number $\nu_r$ also satisfies the Diophantine equation r=$\nu_r p+t_rq$, where $\nu_r$ and $t_r$ are integers, and $0<|\nu_r|<q/2$ \cite{Thouless,dana1985quantised}.

For a quasiperiodic system with an irrational $b$, the spectrum, which is fractal, can be approached by taking an appropriate rational limit with $p,q\rightarrow\infty$. As the system abides by the Diophantine equation for an arbitrarily large $q$, the gaps remain associated with nontrivial Chern numbers. Note that translating the lattice by $m$ sites is equivalent to shifting $\phi$ by $2\pi$($bm$ mod 1). For a rational $b =p/q$, ($bm$ mod 1) has only $q$ different values for all possible translations. However, for a irrational $b$, ($bm$ mod 1) samples the entire $[0,1]$ interval. Thus, the band structure as well as Chern density is guaranteed to be invariant for any shift of $\phi$. Therefore, the Harper models in Eq.~(\ref{harper}) with different $\phi$ have the same bulk band structure and Chern density. 
As a consequence, there is no need for an integration of the Chern density, and the 1D models can be associated with the same Chern number that characterizes the 2D ancestor model. Since in the 2D IQHE, different $b$'s result in different Chern numbers, this indicates that two quasicrystals with different modulation periodicity $b$ generally belong to different topological phases, which cannot be smoothly deformed from one to the other without closing the bulk gap.

Topological states in 1D quasicrystals can also be realized in other 1D models, such as the off-diagonal Harper model, which is defined by setting $\lambda^\mathrm{od} \neq 0, \lambda^\mathrm{d}=0$ and $V_n^{\mathrm{od}}=V_n^\mathrm{H}(\phi)$, and the diagonal/off-diagonal Fibonacci quasicrystal, which is governed by the modulation potential  $V_{n}^{\mathrm{od}} = V_{n}^{\mathrm{F}}=2(\lfloor(n+2) / \tau\rfloor- \lfloor(n+1) / \tau\rfloor)-1 $, where $\tau=(1+\sqrt{5}) / 2$ is the golden ratio, and $\lfloor x\rfloor$ is the floor function. By using the similar dimensional extension procedure, these models are also related to 2D ancestor models. Moreover, as demonstrated in Ref.~\cite{Kraus2}, all these 1D quasicrystal models are topologically equivalent whenever they have the same irrational modulation frequency which corresponds to the same flux quanta per unit cell in the 2D IQHE.

Due to the bulk-boundary correspondence, robust chiral edge states emerge along the edges of 2D IQHEs. The 1D descendant quasicrystal inherits its robust boundary state from the 2D ancestor IQHE. Therefore, by scanning $\phi$ from $-\pi$ to $\pi$, boundary states traverse the gap with $\phi$, as shown in Fig.~\ref{spectrum}. Due to their topological origin, the boundary states are robust against disorder, and cannot be eliminated unless the energy gap closes.

\begin{figure}[hbt]
\centering
\includegraphics[width=0.5\columnwidth]{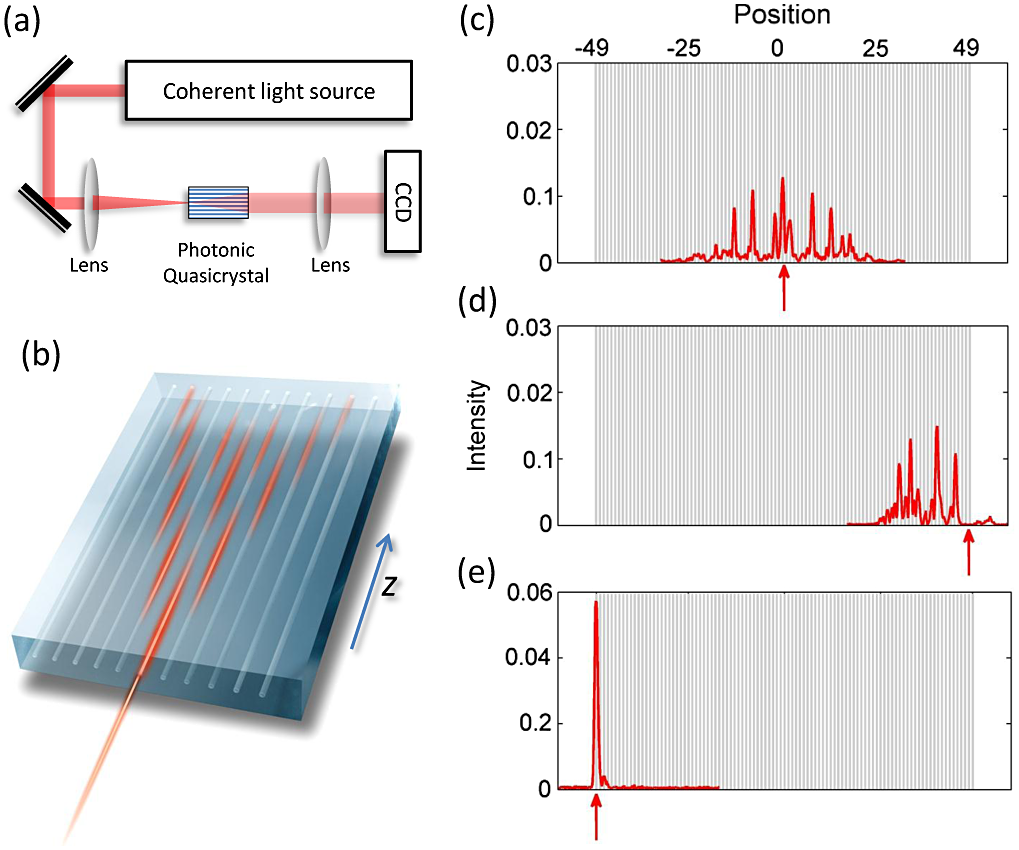}
\caption{(a) A sketch of the experimental setup. (b) An illustration of the conducted experiment. Light is injected into one of the waveguides and tunnels to neighboring waveguides as it propagates. (c)-(e) Experimental observation of the left boundary state for $\phi=$ $\pi / 2 .$ Light was initially injected into a single waveguide (red arrows). The measured outgoing intensity is plotted versus the injection position along the lattice. (c), (d) An excitation at the middle of the lattice (site 0 ) and at the rightmost site (site 49$)$ results in a significant spread. (e) For an excitation at the leftmost site (site -49 ), the light remains tightly localized at the boundary, marking the existence of a boundary state. Reproduced from Ref.~\cite{Kraus}}
\label{photonic QC}
\end{figure}

The 1D quasiperiodic Harper model was experimentally implemented in photonic quasicrystals, which are composed of a 1D quasiperiodic lattice of evanescently coupled single-mode waveguides, as presented in Fig.~\ref{photonic QC}.
The propagation of light in waveguide lattices are described by the discrete nonlinear Schr\"{o}dinger-like equation \cite{christodoulides2003discretizing},
\begin{equation}
\begin{aligned}
i \frac{\partial \psi_{n}}{\partial z}+\left[\beta_{0}+\lambda \cos (2 \pi n \chi)\right] \psi_{n}+& C\left(\psi_{n-1}+\psi_{n+1}\right) \\
&+\gamma\left|\psi_{n}\right|^{2} \psi_{n}=0,
\end{aligned}
\end{equation}
where $\psi_{n}$ is the wave function at site $n$, $z=c t$ is the free propagation axis, $c$ being the speed of light in the medium, $\beta_{0}$ is the single-site eigenvalue of the underlying periodic lattice, $t$ is the hopping amplitude, $\lambda$ controls the on-site amplitude, and $\gamma$ is the Kerr coefficient.
In the linear limit $(\gamma=0)$ the equation is identical to the tight-binding model, with the propagation of light describing the time evolution of the 1D tight-binding model. Modulating the refraction index of the waveguides and the spacing between them controls the on-site and the hopping terms of the Hamiltonian, respectively \cite{christodoulides2003discretizing,szameit2005discrete,lahini2009observation}.

Using this experimental setup, Kraus \textit{et al.} found that light injected in the middle or the rightmost of the lattice showed considerable expansion while light injected in the leftmost of the lattice remained tightly localized, thus signaling the existence of the left edge states, which is consistent with the energy spectrum shown in Fig.~\ref{spectrum}. They further built a similar device in which the waveguides have a slowly varying separation between them along the propagation axis, thus realizing a sweep of $\phi$ in the off-diagonal Harper model. Based on this device, they observed the adiabatic pumping effect in which light injected on one side of the device gradually migrated across the quasicrystal to the other side. Thus, they experimentally demonstrated the topological connection between quasicrystals in 1D and the IQHE in 2D.

The dimensional extension technique is also applicable to quasiperiodic lattices in higher-dimensions, as long as extra degrees of freedom similar to the above $\phi$ can be treated as synthetic dimensions.
It is recently shown that 2D (3D) quasiperiodic structures with additional synthetic dimensions can be mapped to 4D (6D) IQHEs, and these topological phases are characterized by the second (third) Chern number \cite{Kraus3, PhysRevB.98.125431}.

\subsection{Topological phase transitions}
\noindent The topological classification of gapped systems typically assigns a topological integer index (such as the Chern number) to its energy gap \cite{Thouless}, which remains the same as long as the gap does not close. Hence, a system with a given topological index can be continuously deformed into another topologically equivalent system with the same topological index while keeping the bulk gap open. Inversely, when two systems with different topological indices are connected smoothly, the bulk gap must close at the interface between them, which manifests as the appearance of gap-traversing states. Therefore, the existence/absence of interface states and bulk gap closure can be used to determine the topological inequivalence/equivalence between two topological states. As presented in Sec.~\ref{IIIA}, the topological states in 1D quasiperiodic systems can also be assigned Chern numbers which are inherited from their 2D ancestor periodic models, it is, therefore, expected to observe the bulk gap closure when smoothly deforming between topological inequivalent 1D quasicrystals and its absence when the systems are topological equivalent.

\begin{figure}[hbt]
\centering
\includegraphics[width=0.5\columnwidth]{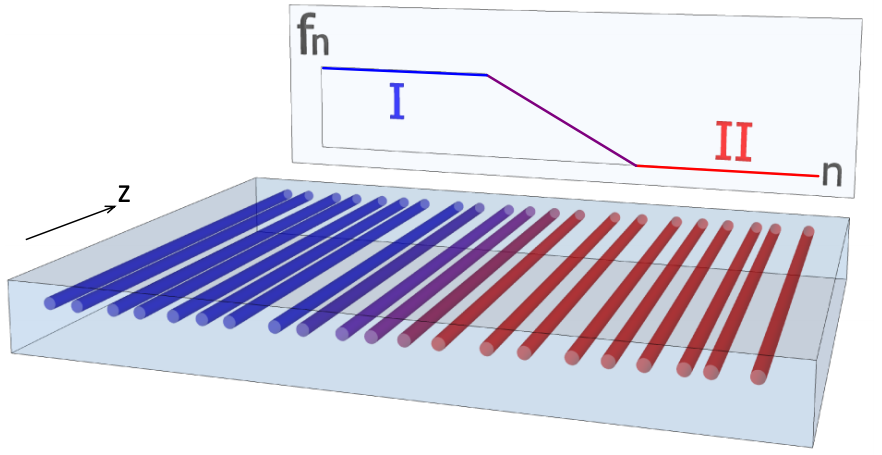}
\caption{Photonic waveguides implementing deformation between two quasicrystals. Reproduced from Ref.~\cite{Verbin}}
\label{deformation}
\end{figure}

In Ref.~\cite{Verbin}, Verbin \textit{et al.} used an array of coupled single-mode waveguides to construct smooth boundaries between topologically distinct or equivalent 1D quasicrystals, as presented in Fig.~\ref{deformation}.
The dynamics of light propagating in these coupled waveguide arrays is described by the tight-binding model, with the propagation axis $z$ taking over the role of time. The general form of Hamiltonian is given by
\begin{equation}
H \psi_n = t_n \psi_{n-1} + t_{n+1} \psi_{n+1},
\end{equation}
where $\psi_n$ is the wave function at waveguide number $n$ and $t_n$ is the hopping amplitude from site $n$ to site $n-1$. As shown in Fig.~\ref{deformation}, the two quasiperiodic systems I and II, where each system has its own set of hopping amplitudes $t_n^\mathrm{I}$ and $t_n^\mathrm{II}$, are smoothly connected by an intermediate region with a deformed hopping profile $t_{n}=f_{n} t_{n}^{\mathrm{I}}+\left(1-f_{n}\right) t_{n}^{\mathrm{II}}$, where
\begin{equation}
f_{n}=\left\{
\begin{array}{ll}
1 & 1 \leq n \leq L_{\mathrm{I}} \\
1-\frac{n-L_{1}}{L_{D}} & L_{\mathrm{I}}<n<L_{\mathrm{I}}+L_{D} \\
0 & L_{\mathrm{I}}+L_{D} \leq n \leq L_{\mathrm{I}}+L_{D}+L_{\mathrm{II}}
\end{array}.\right.
\end{equation}
The length of region I, region II and the deforming region is $L_{\mathrm{I}}$, $L_{\mathrm{II}}$ and $L_D$ respectively.

This interface structure enables the study of the transition between different 1D quasicrystals, on a single waveguide array.
For each quasicrystal, the quasiperiodic hopping amplitude is modulated according to $t_{n}=t_{0}\left[1+\lambda d_{n}\right]$, where $t_0$ is the characteristic hopping amplitude, $\lambda \in [0,1)$ is the modulation strength, and $d_n \in [-1,1]$ is the quasiperiodic modulation function. Specifically, they considered two types of modulations: the Harper modulation
\begin{equation}
d_{n}^{\mathrm{H}}=\cos (2 \pi b n+\phi),
\end{equation}
and the Fibonacci modulation
\begin{equation}
d_{n}^{\mathrm{F}}=2\left(\left\lfloor\frac{\tau}{\tau+1}(n+2)\right\rfloor-\left\lfloor\frac{\tau}{\tau+1}(n+1)\right\rfloor\right)-1=\pm 1,
\end{equation}
where $\tau=(1+\sqrt{5})/2$ is irrational and $\lfloor x\rfloor$ is the floor function. As presented in Sec.~\ref{IIIA}, the Harper quasicrystal is controlled by the irrational modulation frequency $b$, and the Fibonacci-like quasicrystal which is constructed from a sequence of two values as illustrated in Fig.~\ref{projection}. Since the energy spectrum as well as the associated Chern number of the Harper quasicrystal depend on the modulation frequency $b$, two Harper quasicrystals with $b_\mathrm{I}\neq b_{\mathrm{II}}$ are topological inequivalent. Hence, the bulk gap closure is expected at the intermediate region. Comparably, the Fibonacci-like quasicrystal is topologically equivalent with the Harper quasicrystal whenever the modulation frequency $b=(\tau+1)/\tau$ \cite{Kraus2}, and therefore, can be continuously deformed into the Harper quasicrystal without topological phase transition.

\begin{figure*}
\centering
\includegraphics[width=15cm]{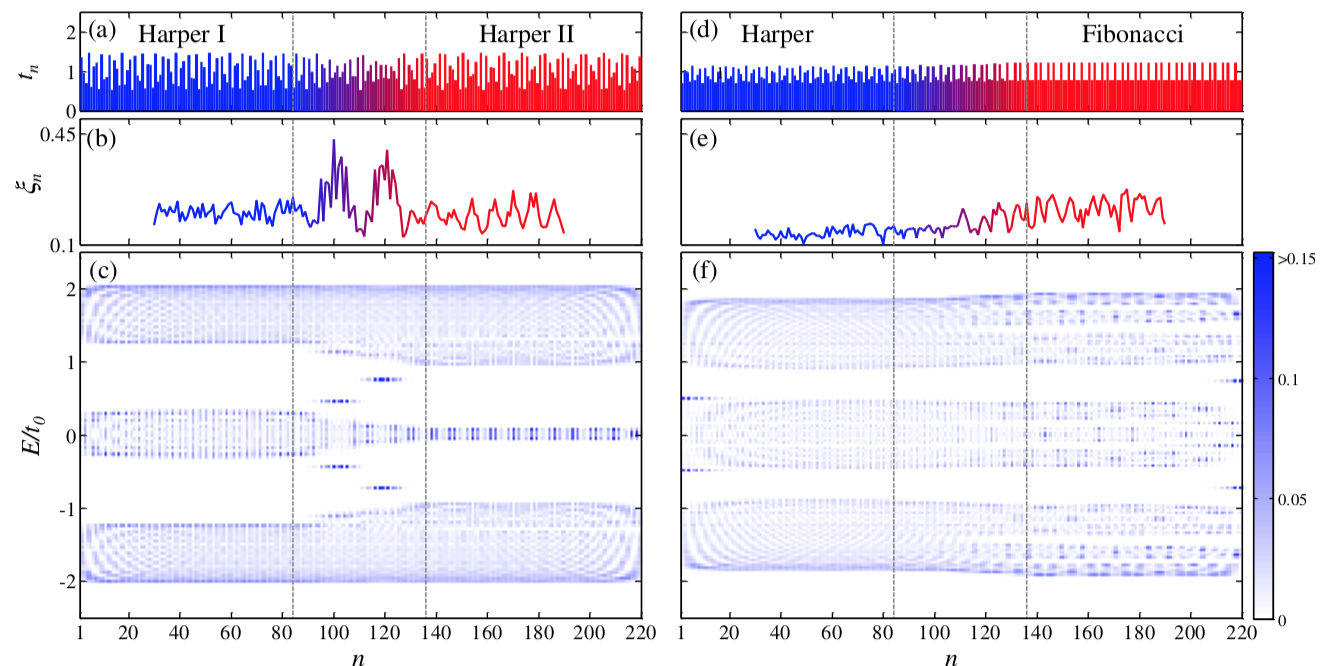}
\caption{Summary of results between (a)-(c) topologically inequivalent Harper model and (d)-(f)topologically equivalent Harper and Fibonacci model. (a),(d) The hopping amplitude $t_n$ as a function of the site index $n$. (b),(e) Experimentally measured return probability as a function of the lattice site $n$. (c),(f) Local density of states (LDOS) of the structure. Reproduced from Ref. \cite{Verbin}}
\label{equiva}
\end{figure*}

To experimentally observe the topological phase transition between the two quasicrystals, they study the continuously deformed photonic quasicrystal by injecting light into one of the waveguides in the array and imaging the outgoing intensity at the output facet using a CCD camera. The width of the outgoing intensity distribution reveals the existence or absence of localized eigenstates near the injection site. To quantify the localization of the outgoing light at different sites, they introduce the generalized return probability
\begin{equation}
\xi = (\sum_{m = n - \Delta}^{n + \Delta} |\psi_m|^2)/(\sum_{m = 1}^{L_{\mathrm{I}} + L_D+L_{\mathrm{II}}} |\psi_m|^2),
\end{equation}
which measures the amount of light that remains within a small distance $\Delta$ from the injection site $n$.

As shown in Fig.~\ref{equiva}(a), two clear peaks appear in the deformation region between topologically inequivalent Harper quasicrystals with modulation frequencies $b_\mathrm{I}\neq b_\mathrm{II}$,  indicating the existence of localized states within the deformation region. These peaks observed in $\xi_n$ are also consistent with the numerically calculated local density of density which describes the spatial distribution of the eigenstates of the structure as a function of energy. On the contrary, no sign of localized states within the deformation region between the topologically equivalent Harper quasicrystal and Fibonacci-like quasicrystal [as shown in Fig.~\ref{equiva}(b)], which agrees with the open gap observed along the deformation in the LDOS. The experimental observation and numerical calculations, which consistently show the absence of phase transition, confirm the topological equivalence between the Fibonacci and the Harper quasicrystals.

\section{Generalization of crystalline topological states in quasicrystals}
\noindent Next, we turn to the realization of quasicrystalline counterparts of topological states already proposed in crystals. As various 2D topological states have been discovered in crystals, it is natural to generalize these states to 2D quasicrystals. Since quasicrystals possess long-range orientational order but lack translational symmetry, one cannot use the Bloch theorem as for crystals. Moreover, various topological invariants defined for periodic systems, are no longer applicable in quasicrystals.
Here we mainly discuss basic tight-binding Hamiltonians to realize 2D topological states in 2D quasicrystals and topological invariants to identify their topological nature.

\subsection{Quantum Hall states}
\begin{figure}[hbt]
\centering
\includegraphics[width=6cm]{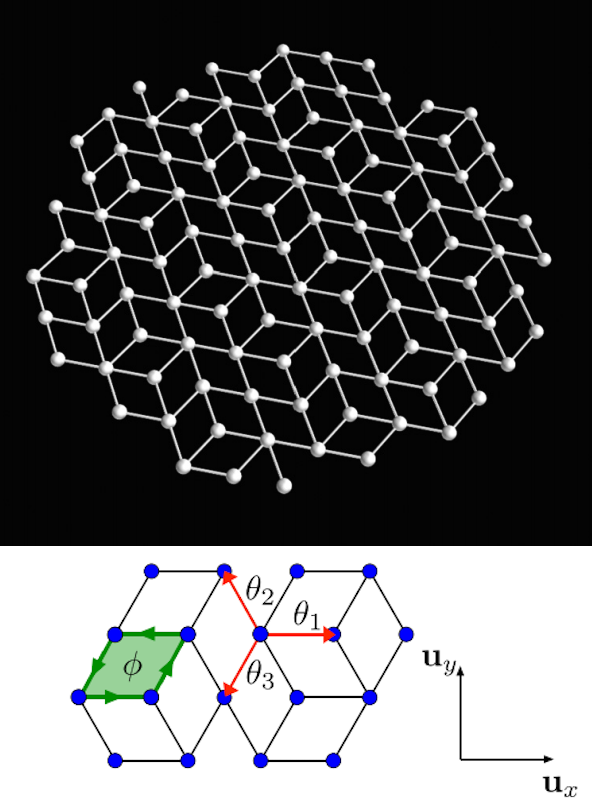}
\caption{Top: Isometric generalized Rauzy tiling. Bottom: The red arrows represent directions of the three nearest hoppings, their corresponding Peierls phase factors are shown. The magnetic flux per tile is $\phi=B l^{2} \sqrt{3} / 2$.
Reproduced from Ref.~\cite{Tran}}
\label{rauzy}
\end{figure}

\noindent The IQHE is a well-known example of 2D topological states \cite{Klitzing}. In an IQHE, 2D lattices subjected to a strong perpendicular magnetic field display nontrivial topological bands characterized by the Chern number \cite{Thouless}. As the 2D lattice is not necessarily periodic, it is possible to generalize the IQHE to 2D quasicrystals. There has been some research exploring the IQHE in 2D quasicrystals \cite{Tran,Hofstadter2,Hofstadter3,PhysRevB.101.115413}. For example, Tran \textit{et al.}\cite{Tran} investigate the topological properties of a 2D quasicrystal subjected to a uniform magnetic field. The 2D quasicrystal is based on the generalized Rauzy tilling, as presented in Fig.~\ref{rauzy}.

In order to describe the quasicrystal in the presence of a uniform magnetic field, they introduce the spinless tight-binding model with Peierls phase factors:
\begin{equation}
\hat{H} = -J \sum_{\langle j,k\rangle} e^{i \theta_{jk}} \hat{c}_k^{\dagger} \hat{c}_j, \quad \theta_{jk} = \int_{\boldsymbol{r}_j}^{\boldsymbol{r}_k} \mathbf{A} \cdot d\mathbf{l},
\label{Hbutterfly}
\end{equation}
where $c_{j}^{\dagger}$ creates a fermion at the lattice site $\boldsymbol{r}_{j}, J$ is the hopping matrix element, $\exp \left(i \theta_{j k}\right)$ denotes the Peierls phase factor due to the magnetic field \cite{Hofstadter}, and $\mathbf{A}$ is the corresponding vector potential. The magnetic flux per tile is $\phi=Bl^2\sqrt{3}/2$ over the entire quasicrystal, where $l$ is the distance between neighboring sites. The flux quantum equals $\phi_0 = 2\pi$ in the present units where $\hbar=e=1$.
\begin{figure}[hbt]
\centering
\includegraphics[width=0.5\columnwidth]{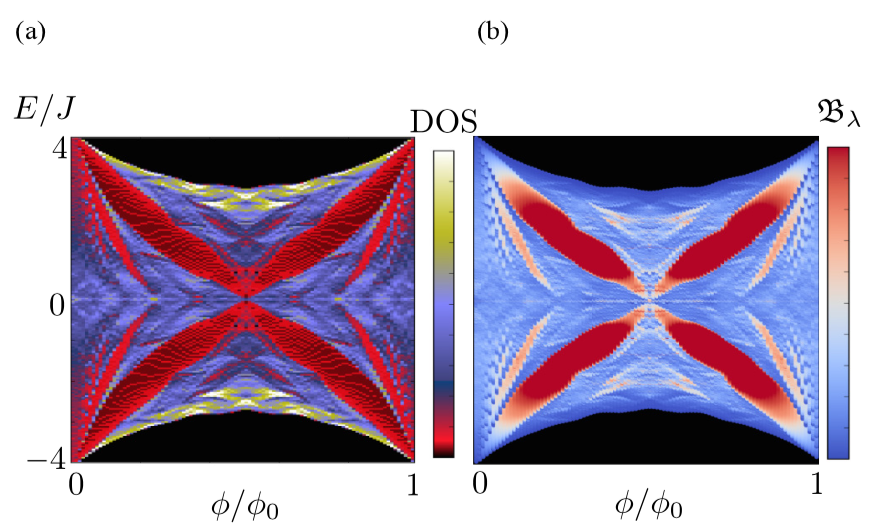}
\caption{(a) Density of states (DOS) in terms of the magnetic flux $\phi$ for the Rauzy tiling quasicrystal with an OBC. (b) The edge-locality marker $\mathfrak{B}_{\lambda}$ as a function of the flux $\phi$ and eigenenergy $E_{\lambda}$. The states lying in the low-DOS regions of (a) are located at the boundary. Reproduced from Ref.~\cite{Tran}}
\label{butterfly}
\end{figure}
The density of states (DOS) of the Rauzy tiling with open boundary conditions (OBCs) were calculated based on Eq.~(\ref{Hbutterfly}). Figure~\ref{butterfly}(a) depicts the DOS in terms of the magnetic flux $\phi$, which shows similarities with the Hofstadter's butterfly in the square lattices case \cite{Hofstadter}. Moreover, this figure has regions of very low density (in red), which correspond to the spectral gaps in the closed (torus) geometry (i.e., in a PBC). To further analyze the properties of the DOS diagram, they used the edge-locality marker
\begin{equation}
\mathcal{B}_{\lambda} = \sum_{\mathbf{r} \in \mathrm{edge}} |\psi_{\lambda} (\mathbf{r})|^2,
\end{equation}
to characterize the localization of each eigenstate $\psi_{\lambda}$, with eigenenergy $E_{\lambda}$, at the boundary. As shown in Fig.~\ref{butterfly}(b), the distribution of $\mathcal{B}_{\lambda}$ over all states in the $E_{\lambda}$-$\phi$ plane follows the shape of the butterfly in Fig.~\ref{butterfly}(a), indicating that these low-DOS regions host chiral edge states. This is in agreement with the fact that these regions correspond to bulk energy gaps in the closed geometry.

As in periodic lattices, the chiral states residing within the low-DOS regions of the spectrum are associated with a Chern number. However, the conventional numerical method of computing the Chern number \cite{doi:10.1143/JPSJ.74.1674,hatsugai2006topological} relies on the restoration of spatial periodicity. Due to the lack of translational symmetry, it is more satisfactory to evaluate the topological number without invoking any reciprocal-space parameters (such as the wave vector $k$). This is achieved using the \textit{real-space Chern invariant} $\mathcal{C}$, as introduced by Bianco and Resta \cite{Bianco}. This method provides a local characterization of the bulk topological invariant, and hence, is independent of the boundary conditions.

As introduced by Bianco and Resta \cite{Bianco}, the real-space Chern invariant can characterize the topology of finite-size systems with OBC locally in real space. Tran \textit{et al.} \cite{Tran} apply this method to quasicrystals.
To do so, they introduce the local Chern marker
\begin{equation}
\mathfrak{C}\left(\mathbf{r}_{i}\right)=-4 \pi \operatorname{Im}\left[\sum_{\mathbf{r}_{j}}\left\langle\mathbf{r}_{i}\left|\hat{x}_{\mathcal{Q}}\right| \mathbf{r}_{j}\right\rangle\left\langle\mathbf{r}_{j}\left|\hat{y}_{\mathcal{P}}\right| \mathbf{r}_{i}\right\rangle\right],
\end{equation}
where
\begin{eqnarray}
\left\langle\mathbf{r}_{i}\left|\hat{x}_{\mathcal{Q}}\right| \mathbf{r}_{j}\right\rangle&=&\sum_{\mathbf{r}_{k}} \mathcal{Q}\left(\mathbf{r}_{i}, \mathbf{r}_{k}\right) x_{k} \mathcal{P}\left(\mathbf{r}_{k}, \mathbf{r}_{j}\right),\\
\left\langle\mathbf{r}_{j}\left|\hat{y}_{\mathcal{P}}\right| \mathbf{r}_{i}\right\rangle&=&\sum_{\mathbf{r}_{k}} \mathcal{P}\left(\mathbf{r}_{j}, \mathbf{r}_{k}\right) y_{k} \mathcal{Q}\left(\mathbf{r}_{k}, \mathbf{r}_{i}\right),
\end{eqnarray}
and the projection operator in the position basis is defined as
\begin{eqnarray}
\mathcal{P}\left(\mathbf{r}_{i}, \mathbf{r}_{j}\right)&=&\sum_{E_{\lambda}<E_{F}}\left\langle\mathbf{r}_{i} \mid \psi_{\lambda}\right\rangle\left\langle\psi_{\lambda} \mid \mathbf{r}_{j}\right\rangle,\\
\mathcal{Q}\left(\mathbf{r}_{i}, \mathbf{r}_{j}\right)&=&\sum_{E_{\lambda}>E_{F}}\left\langle\mathbf{r}_{i} \mid \psi_{\lambda}\right\rangle\left\langle\psi_{\lambda} \mid \mathbf{r}_{j}\right\rangle,
\end{eqnarray}
where $\left\{\left|\mathbf{r}_{i}\right\rangle\right\}$ denotes the lattice-position basis.
For a nonperiodic system, a smoothened real-space Chern number is obtained by an average of $\mathfrak{C}(\mathbf{r}_{i})$ over a disk $D$ of radius $r_D$, cebtered ariybd $\mathbf{r}_0$ and located within the bulk,
\begin{equation}
\mathcal{C}_{D}\left(\mathbf{r}_{0}\right)=\frac{1}{A_{D}} \int_{D} \mathfrak{C}\left(\mathbf{r}^{\prime}\right) d \mathbf{r}^{\prime},
\label{localChern}
\end{equation}
where $A_D$ is the disk area.  Note that the quantity $\mathcal{C}_{D}\left(\mathbf{r}_{0}\right)$ in Eq.~(\ref{localChern}) is still local, and thus, it can be exploited to probe (potentially different) topological orders in nonperiodic systems.

For the specific Rauzy quasicrystal studied by Tran \textit{et al.}, the local real-space Chern number is obtained by the averaged marker $\mathfrak{C}(\mathbf{r}_{i})$ over a single quasicrystal tile
\begin{equation}
\mathcal{C}\left(\mathbf{r}_{j}\right)=\frac{\mathfrak{C}\left(\mathbf{r}_{j}\right)}{A_{\mathrm{tile}}}, \quad A_{\mathrm{tile}}=l^{2} \sqrt{3} / 2.
\end{equation}
Then, to reduce the fluctuations of the local real-space Chern number in the quasicrystal, the smoothened real-space Chern number is calculated as
\begin{equation}
\mathcal{C}_{D}\left(\mathbf{r}_{0}\right)=\frac{1}{N} \sum_{j \in D} \mathcal{C}\left(\mathbf{r}_{j}\right),
\end{equation}
where the average is performed over the $N$ points within the disk $D$. The calculated real-space Chern numbers of the Rauzy quasicrystal with different magnetic flux per tile are all integer within small errors, indicating that the bulk topology of quasicrystals subjected to a uniform magnetic field is well captured by the (local) real-space Chern number.

Topological states appearing in electronic systems can usually be generalized to photonic systems \cite{floquet1,floquet2,lindner2011floquet}. According to Eq.~(\ref{Hphotonic}), the paraxial wave equation which describes the propagation of optical wave in photonic lattices, is mathematically equivalent to the Schr\"{o}dinger equation---with the propagation coordinate $z$ playing the role of time, and the local change in the refractive index being the potential. As an analogue of IQHE in 2D quasicrystals, Bandres \textit{et al.}\cite{Bandres} proposed a topological Floquet photonic quasicrystals in a 2D photonic quasicrystalline lattice composed of evanescently coupled helical waveguides. In this system, the role of external magnetic fields as for IQHE is replaced by an artificial gauge field via dynamic modulation.

\begin{figure}[hbt]
\centering
\includegraphics[width=0.5\columnwidth]{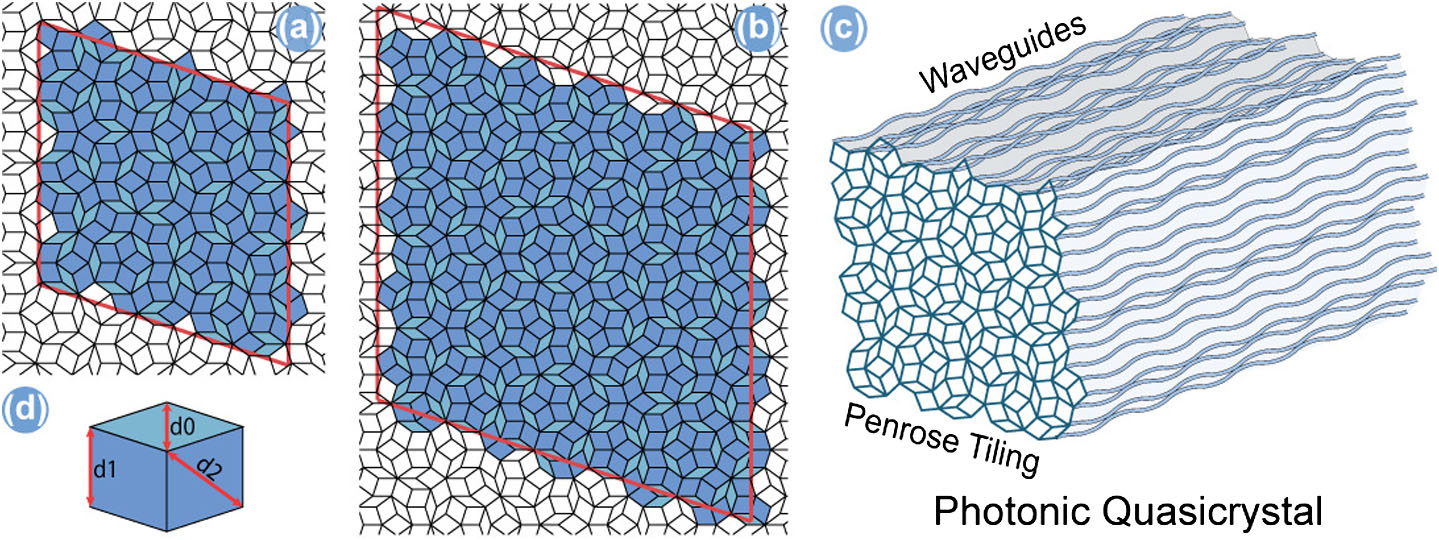}
\caption{(a),(b) Periodic Penrose approximants of Penrose quasicrystals, containing 199 and 521 vertices. (c) Photonic Floquet quasicrystal formed by introducing a helical waveguide in each vertex of the Penrose tilling lattice. Reproduced from Ref. ~\cite{Bandres}}
\label{floquet1}
\end{figure}

Specifically, they consider a Penrose tilling quasicrystalline lattice with helical waveguides attached to each vertex, as shown in Fig.~\ref{floquet1}. Similar to the IQHE, the system can be described by a tight-binding Hamiltonian subject to a vector potential given by
\begin{equation}
i \partial_z \Psi_n = \sum_{\langle m \rangle} c_{mn} e^{iA(z) \cdot r_{mn}} \Psi_m,
\end{equation}
where $\Psi_n$ is the amplitude in the n-the site, $c_{mn}$ and $r_{mn}$ are the coupling constant and the displacement between site $m$ and $n$, respectively. $A(z) = A_0[\cos(\Omega z),\sin(\Omega z)]$ is the vector potential (reflecting the helicity of the waveguides). As the Hamiltonian is $Z=2\pi/\Omega$ periodic, the Floquet theory applies \cite{gu2011floquet}, and the Floquet eigenmodes are of the form $\Psi_n(z)=\exp(i\beta z)\phi_n(z)$, where $\phi_n(z)$ is $z$ periodic and $\beta$ is the Floquet eigenvalue, or ``quasienergy'', which are defined modulo the frequency $\Omega$. The existence of the topological state is demonstrated by directly calculating the Bott index, which is another topological index equivalent to the Chern number \cite{toniolo2017equivalence}, and by studying the unidirectional transport of the gapless edge states and its robustness in the presence of defects.

\subsection{Quantum anomalous Hall states}
\noindent After the discovery of the IQHE, Haldane proposed a model (termed the Haldane model) on a honeycomb lattice in which an IQHE appears as an intrinsic property of its topological band structure, rather than being induced by a strong external magnetic field \cite{Haldane}. Such a quantum Hall effect without magnetic fields and associated Landau levels is also known as the ``quantum'' version of anomalous Hall effect, which is named as the quantum anomalous Hall (QAH) state or the Chern insulator (CI).
Recently, QAH states were realized in quasicrystalline systems, such as quasicrystalline models with staggered flux \cite{He} and 30° twisted bilayer graphene \cite{li2020quantum,yao2018quasicrystalline}.

\begin{figure}[hbt]
\centering
\includegraphics[width=6cm]{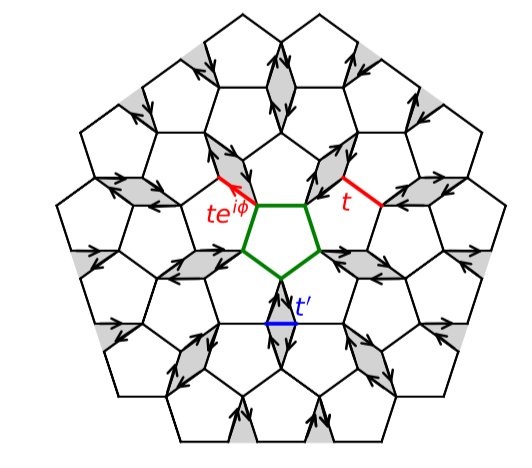}
\caption{Type-I quasicrystalline CI model realized by introducing staggered flux in D\"{u}rer's tiling quasicrystal lattice. Reproduced from Ref.~\cite{He}}
\label{QAH1}
\end{figure}

\begin{figure}[hbt]
\centering
\includegraphics[width=0.5\columnwidth]{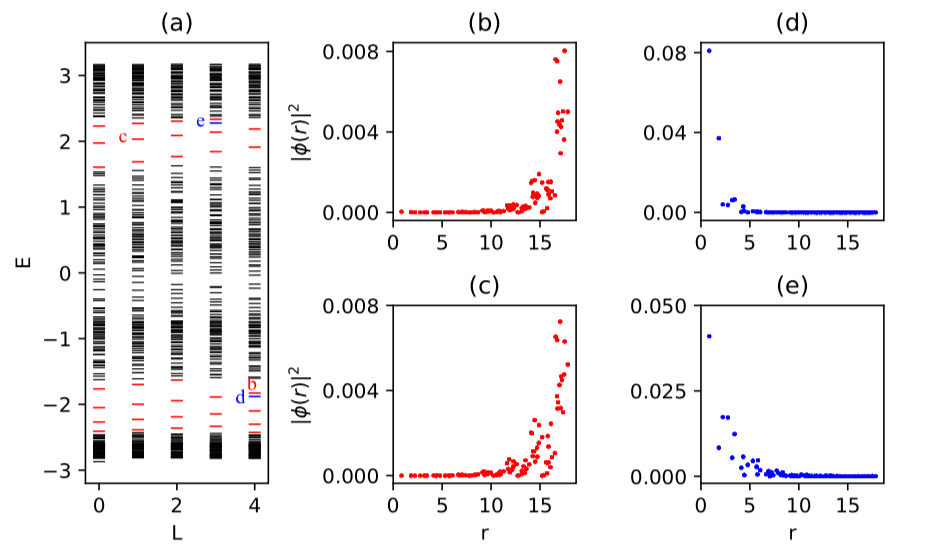}
\caption{(a) Energy spectrum of the Type-I quasicrystalline CI model. L is the quantum number of the angular momentum. (b) and (c) are the spatial distributions of wave function $|\psi(r)|^{2}$ for edge states highlighted in (a). (d) and (e) are similar $|\psi(r)|^{2}$ but for the correspondingly core states in (a). Reproduced from Ref. \cite{He}}
\label{QAH2}
\end{figure}

As an example, here we introduce a  quasicrystalline CI \cite{He}, which is constructed by elaborately imposing staggered flux on disk geometry with fivefold rotational symmetry in D\"{u}rer's pentagonal quasicrystal, as shown in Fig.~\ref{QAH1}.
Similar to the Haldane model, an inequivalent staggered flux is introduced on the polygons in the quasicrystalline lattice, which results in different types of quasicrystalline CIs according to the distribution of the staggered fluxes. Taking the type-I model as an example (see Fig.~\ref{QAH1}), the staggered fluxes of $-4\phi$ are imposed on all diamonds (shadow), and $+2\phi$ for all the pentagons (bright) except the central pentagon.
The total flux in the whole disk is exactly zero if the regular pentagon shape is considered.
The real-space Hamiltonian is given by
\begin{equation}
 H  = -t \sum_{\langle \mathbf{r}\mathbf{r}' \rangle} a^{\dagger}_{\mathbf{r}'} a_{\mathbf{r}} e^{i \phi_{\mathbf{r}\mathbf{r}'}} -t' \sum_{\Diamond,\langle \mathbf{r}\mathbf{r}'\rangle} a^{\dagger}_{\mathbf{r}'} a_{\mathbf{r}},
\end{equation}
where $a_{\mathbf{r}}^{\dagger}\left(a_{\mathbf{r}}\right)$ creates (annihilates) a particle at vertex (site) $\mathbf{r}$, $t$ (which is set as unit) and $t'$ are nearest-neighbor (NN) and next-nearest-neighbor (NNN) hopping, respectively. $\left\langle\mathbf{r r}^{\prime}\right\rangle$ runs over all the NN sites, and $\Diamond,\left\langle\mathbf{r r}^{\prime}\right\rangle^{\prime}$ denotes the NNN sites in each diamond. $\phi_{\mathbf{r}^{\prime} \mathbf{r}}$ is the phase difference between the NN sites as shown in Fig.~\ref{QAH1}.
The topological feature of this model is investigated by directly calculating the energy spectrum where edge states are observed in the bulk gap, and the spatial distribution of edge states, which are mainly localized near the boundaries, as shown in Fig.~\ref{QAH2}.

\begin{figure}[hbt]
\centering
\includegraphics[width=8cm]{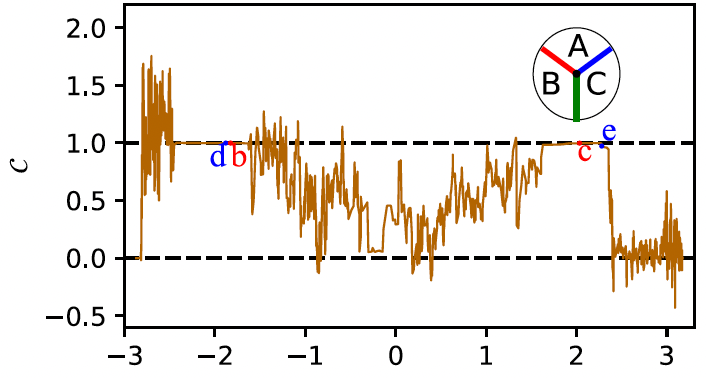}
\caption{Real-space Chern number as a function of Fermi energy. Reproduced from Ref.~\cite{He}}
\label{QAH3}
\end{figure}

To further identify the quasicrystalline CI, they calculated the Chern number using the Kitaev formula and the local Chern number marker. Here we present the method from the \textit{Kitaev formula} \cite{kitaev2006QAH}, which defines the real-space Chern number as
\begin{equation}
\mathcal{C} = 12\pi i \sum_{j\in A} \sum_{k\in B} \sum_{l\in C}(P_{jk} P_{kl} P_{lj} - P_{jl} P_{lk} P_{kj}).
\end{equation}
The disk of the quasicrystal is now cut into three distinct neighboring regions (labelled as A, B and C) arranged in the counterclockwise order, as shown in insert in Fig.~\ref{QAH3}. $j, k,$ and $l$ denote the vertex (or site) in three regions, respectively. $\hat{P}=\sum_{E_{n}<E_{F}}\left|\phi_{n}\right\rangle\left\langle\phi_{n}\right|$ is the projection operator of occupied states below the Fermi energy $E_{F}$, and $P_{j k}=\sum_{E_{n}<E_{F}} \phi_{n}\left(r_{j}\right) \phi_{n}\left(r_{k}\right)^{*}$ is the matrix elements of $\hat{P}$ with $\phi_{n}\left(r_{j}\right)=\left\langle j \mid \phi_{n}\right\rangle$. The real-space Chern number is independent of the choices of the A, B and C regions \cite{real_CN}. As the Chern number is determined by all occupied states, it is apparently related to the position of the Fermi level.
As shown in Fig.~\ref{QAH3}, the real-space Chern number shows a quantized plateau when the Fermi energy is located in the bulk gap, indicating the nontrivial topology of the quasicrystal.

\subsection{Quantum spin Hall states}
\noindent As a cousin of the IQHE, the quantum spin Hall (QSH) effect which exhibits a quantized spin-Hall conductance 
and a vanishing charge-Hall conductance, have attracted considerable attention since it was initially proposed by Kane and Mele \cite{kane2005quantum}, and independently by Bernevig and Zhang \cite{PhysRevLett.96.106802}.
Different from the IQHE which is associated with a nonzero Chern number, QSH states are characterized by the topological invariant called the $\mathbb{Z}_2$ index, which is zero for trivial and one for nontrivial cases \cite{kane2005z}. Moreover, the nontrivial topological nature of the QSH state also manifests as topologically protected metallic edge states with helical spin polarization residing in an insulating bulk gap. For comprehensive reviews of QSH states please see Refs.
\cite{maciejko2011quantum,konig2008quantum,qi2011topological,RevModPhys.82.3045} and references therein.

\begin{figure}[hbt]
\centering
\includegraphics[width=6cm]{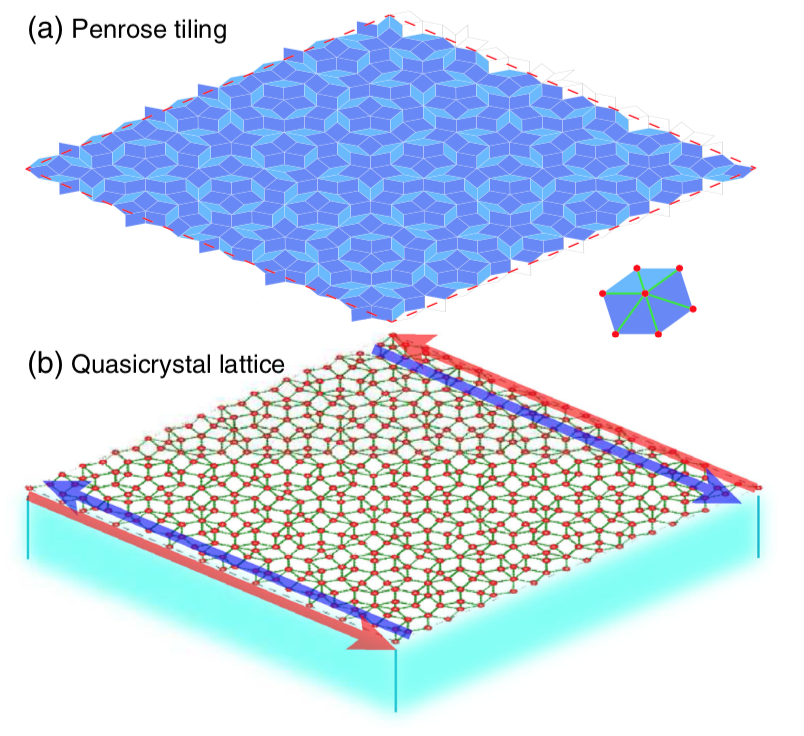}
\caption{(a) Penrose tiling pattern containing 521 vertices. (b) Atomic model of a QSH state in a surface-based 2D quasicrystal. The red and blue arrows represent helical edge states with opposite spin polarizations. Reproduced from Ref.~\cite{Huang}}
\label{penrose}
\end{figure}

Recently, the QSH effect is also explored in quasicrystals \cite{Huang,Huang1,Huang3} by one of the authors. As shown in Fig.~\ref{penrose}, Huang and Liu \cite{Huang} map a 2D Penrose tiling to an atomic quasicrystalline lattice with atomic orbitals, which is makeable by experiments.
We consider a general atomic basis tight-binding model for quasicrystalline lattices with three orbitals ($s, p_x,p_y$) per site, which is given by
\begin{equation}
\begin{split}
 H  = \sum_{i \alpha} \epsilon_{\alpha} c^{\dagger}_{i\alpha} c_{i \alpha} + \sum_{\langle i \alpha, j \beta \rangle} t_{i\alpha ,j \beta}  c^{\dagger}_{i\alpha} c_{j \beta} \\
+ i\lambda \sum_{i} (c^{\dagger}_{i p_y} \sigma_z c_{i p_x} - c^{\dagger}_{i p_x} \sigma_z c_{i p_y}),
\label{Hqsh}
\end{split}
\end{equation}
where $c_{i \alpha}^{\dagger}=\left(c_{i \alpha \uparrow}^{\dagger}, c_{i \alpha \downarrow}^{\dagger}\right)$ are electron creation operators on
the $\alpha\left(=s, p_{x}, p_{y}\right)$ orbital at the $i$ th site. $\epsilon_{\alpha}$ is the on-site energy of the $\alpha$ orbital. The second term is the hopping term, where $t_{i \alpha, j \beta}=t_{\alpha, \beta}\left(\mathbf{d}_{i j}\right)$ is the hopping integral, which depends on the orbital type $(\alpha$ and $\beta)$ and the vector $\mathbf{d}_{i j}$ between sites $i$ and $j$. $\lambda$ is the spin-orbit coupling $(\mathrm{SOC})$ strength. In our model, the hopping integral follows the Slater-Koster formula \cite{SK}
\begin{equation}
t_{\alpha, \beta}\left(\mathbf{d}_{i j}\right)=\operatorname{SK}\left[V_{\alpha \beta}\left(d_{i j}\right), \hat{\mathbf{d}}_{i j}\right],
\end{equation}
where $\hat{\mathbf{d}}_{i j}$ is the unit direction vector. The distance dependence of the bonding parameters $V_{\alpha \beta}\left(d_{i j}\right)$ is captured approximately by the Harrison relation \cite{Harrison}:
\begin{equation}
V_{\alpha \beta}\left(d_{i j}\right)=V_{\alpha \beta, 0} \frac{d_{0}^{2}}{d_{i j}^{2}}.
\end{equation}
where $d_0$ is a scaling factor to uniformly tune the bonding strengths.

\begin{figure*}[hbt]
\centering
\includegraphics[width=15cm]{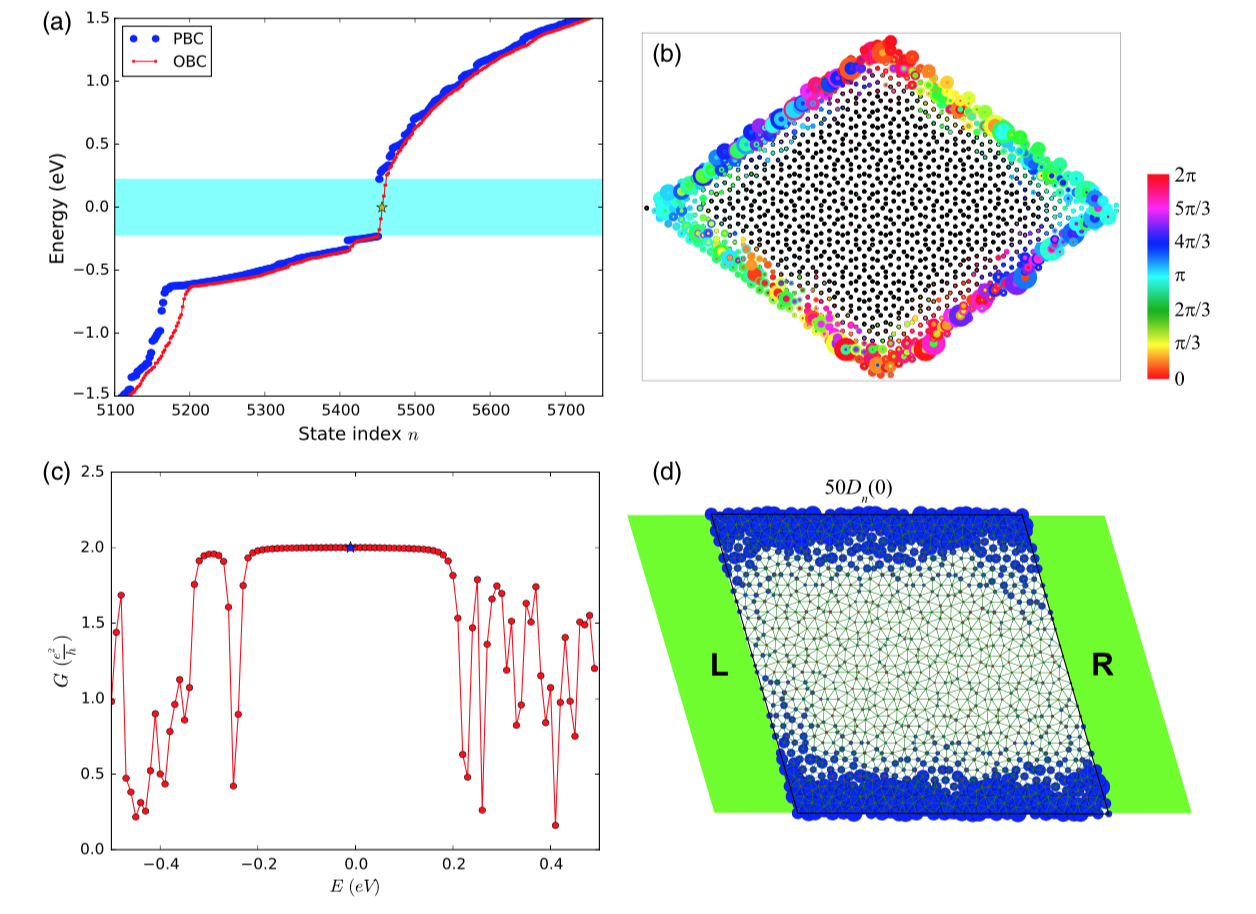}
\caption{Calculation of a Penrose-type quasicrystal sample with 1364 atoms. (a) Energy eigenvalues $E_n$ versus the state index $n$. The system with a PBC shows a gap, while that with an OBC shows midgap states. (b) The wave function $|\psi(\mathbf{r})\rangle= \rho(\mathbf{r})e^{i\phi(\mathbf{r})}$ of the midgap state [marked as the yellow star in (a)] is localized on the edge of the system. The size and the color of the blob indicate the norm $|\rho(\mathbf{r})|^2$ and phase $\phi(\mathbf{r})$ of the wave function, respectively. (c) Two-terminal conductance $G$ as a function of the Fermi energy $E$, showing a quantized plateau in the energy gap. (d) Local density of state $D_n(E)$ at $E =0$ eV for the central quasicrystal in the transport simulation. The size of the blue dot represents the relative value of the local density of state. Reproduced with permission from Ref.~\cite{Huang}}
\label{transport}
\end{figure*}

With proper tight-binding parameters, a band inversion between \textit{s} and \textit{p} states of different parities can be realized in the model, which gives rise to a QSH state in the quasicrystal. As shown in Fig.~\ref{transport}(a), the energy spectrum of the quasicrystal approximant with an artificial PBC has an energy gap around the Fermi level, while some edge states which are localized on the boundary of the finite sample [see Fig.~\ref{transport}(b)], emerge in the gap in the presence of an OBC, implying the existence of nontrivial topology. The conductive feature of the edge states is verified by a quantum transport simulation based on the nonequilibrium Green's function method. Remarkably, a clear quantized plateau at $G=2e^2/h$ for the two-terminal charge conductance is observed, and the conductive channels are mostly contributed by the topological edge states, as seen from the local density of states in Fig.~\ref{transport}(d).

To further identify the QSH states in quasicrystals, we define the \textit{spin Bott index} $B_s$, which is topologically equivalent to the $\mathbb{Z}_2$ index, for QSH states in nonperiodic systems. The inspiration was drawn from the equivalence between the Bott index and the Chern number
\cite{toniolo2017equivalence}, and the definition of the spin Chern number for QSH states in crystals \cite{spin_Chern,spin_Chern2,spin_Chern3}. 
Although the lack of periodicity hinders the usage of crystalline momentum, we can still obtain the eigenenergies and eigenstates of the Hamiltonian in quasicrystals via a direct diagonalization, and construct the projector operator of the occupied states $P$ and the projected spin operator,
\begin{equation}
P = \sum_i^{N_{occ}} |\psi_i \rangle \langle \psi_i |,\quad P_z = P \hat{s_z} P,
\label{projector}
\end{equation}
Then, we make a smooth decomposition by solving the eigenvalue problem
\begin{equation}
P_z|\phi_\pm\rangle=S_\pm|\phi_\pm\rangle,
\end{equation}
and construct new projector operators for two spin sectors,
\begin{equation}
P_{\pm} = \sum_i^{N_{occ}/2}  |\pm \phi_i\rangle \langle\pm \phi_i|.
\end{equation}
Next, we calculate the projected position operators
\begin{equation}
\begin{split}
U_{\pm} = P_{\pm} e^{i2\pi X} P_{\pm} + (I - P_{\pm}),
\\
V_{\pm} = P_{\pm} e^{i2\pi Y} P_{\pm} + (I - P_{\pm}),
\end{split}
\label{xy}
\end{equation}
where $X$ and $Y$ are the rescaled coordinates which are defined in the interval [0,1). For each spin sector, the Bott index, which measures the commutativity of the projected position operators \cite{Bott1,Bott2,Bott3,Bott4}, is given by
\begin{equation}
B_\pm=\frac{1}{2\pi}\textrm{Im}\{\textrm{Tr}[\log({V}_\pm {U}_\pm V_\pm^\dag U_\pm^\dag)]\}.
\end{equation}
Finally, the spin Bott index is defined as the half difference between the Bott indices for the two spin sectors.
\begin{equation}
B_s = \frac{1}{2} (B_{+} - B_{-}).
\label{Bs}
\end{equation}
The spin Bott index is a well-defined topological invariant, which applies to quasiperiodic and amorphous systems. Therefore, it provides a useful tool to determine the electronic topology of those systems without translational symmetry. For the quasicrystal in Fig.~\ref{penrose} the calculated spin Bott index $B_s = 1$, indicating indeed a QSH state.

Following our work, Chen \textit{et al.} studied the effect of disorder on the QSH state in quasicrystals and found that disorder-induced topological Anderson insulators can also be realized in quasicrystals \cite{TAI1,TAI2}. Based on the model in Eq.~(\ref{Hqsh}), they consider random on-site disorder on the Penrose-type quasicrystal. It is found that a disorder can induce a phase transition from a normal insulator to a QSH state in the quasicrystal system, indicating the emergence of a topological Anderson insulator state.
The transport simulation shows that a quantized two-terminal conductance plateau can arise inside the energy gap of the normal-insulator phase for moderate Anderson disorder strength. This topological Anderson insulator is further identified by the disorder-averaged spin Bott index.


\subsection{Topological crystalline insulators}
\noindent The discovery of QSH states in quasicrystals stimulates us to investigate other symmetry-protected topological states in quasicrystals. TCIs are special states of matter in which the topological nature of electronic structures arises from crystal symmetries, such as mirror or rotation symmetries. And a key characteristic of TCIs is the presence of metallic boundary states on surfaces/edges preserving that symmetry.
Since TCIs could be induced by a band inversion between states with the same parity but different eigenvalues of some lattice symmetries, we explore the realization of TCIs in quasicrystals which stem from the band inversion mechanism. In Ref.~\cite{Huang2}, we propose the concept of aperiodic TCIs, as exemplified by an octagonal quasicrystal.

\begin{figure}[hbt]
\centering
\includegraphics[width=0.5\columnwidth ]{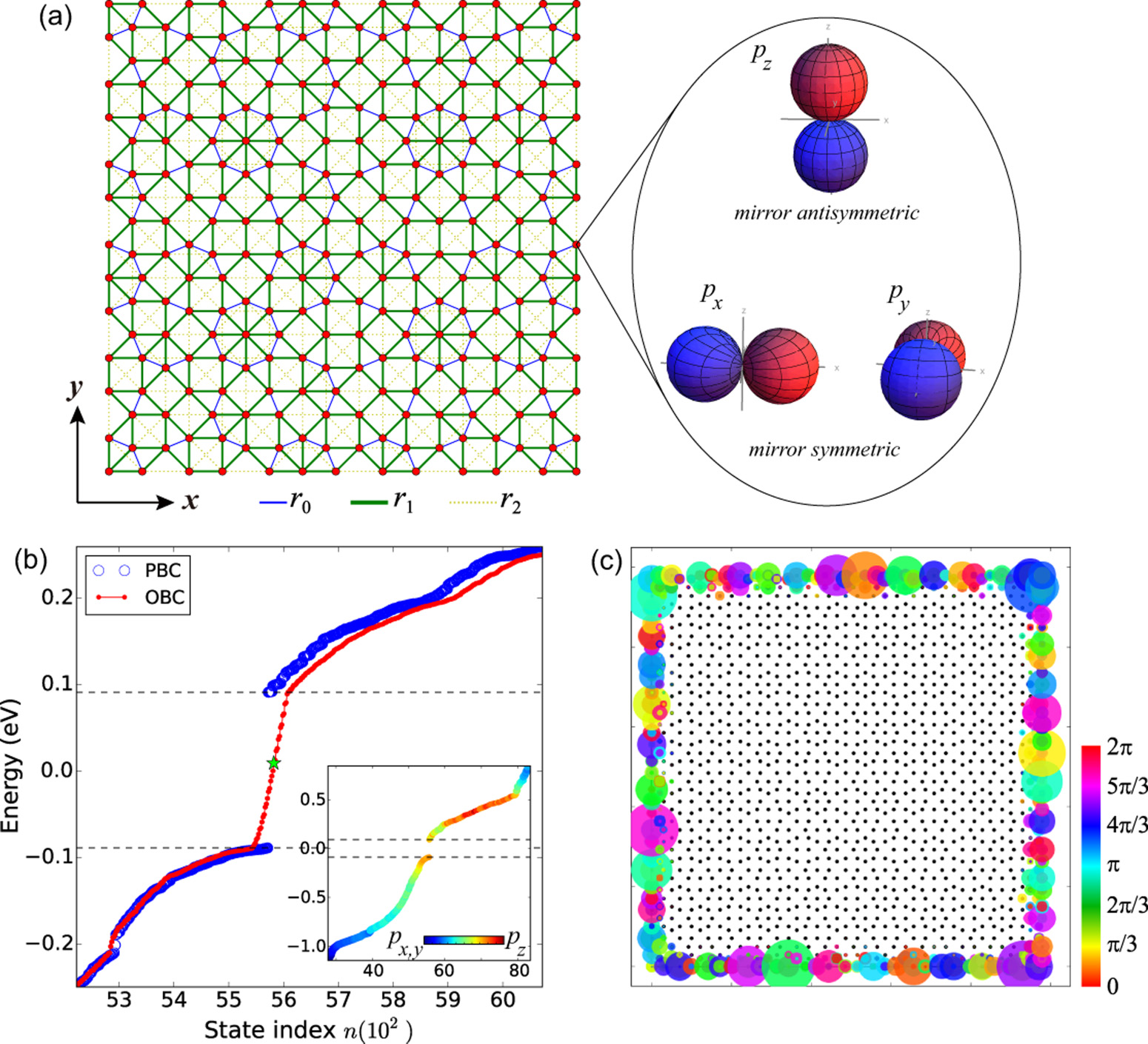}
\caption{(a) Octagonal quasicrystalline lattice based on the Ammann-Beenker tiling containing 264 vertices. Three atomic p orbitals with different mirror symmetries are placed on vertices. (b) Energy eigenvalues
$E_n$ versus the state index $n$ from the calculation of a quasicrystal sample with 1393 atoms. The inset shows the orbital-resolved spectrum of the quasicrystal with a PBC. (c) The wave function 
of the midgap state [marked as the green star in (a)] is located on the edge. 
Reproduced from Ref.~\cite{Huang2}}
\label{TCI}
\end{figure}

To demonstrate the realization of aperiodic TCIs, we use a generic atomic-basis model on a 2D quasicrystals according to the octagonal Ammann-Beenker (AB) tiling, as shown in Fig.~\ref{TCI}(a). 
There are three orbitals ($p_x, p_y$, and $p_z$) per site having opposite eigenvalues with respect to the mirror operation $\hat{M}_z$. The Hamiltonian is expressed as
\begin{equation}
\begin{split}
 H  = & \sum_{i \alpha \mu} \epsilon_{\alpha} c^{\dagger}_{i\alpha\mu} c_{i \alpha \mu} + \sum_{\langle i \alpha, j \beta \rangle,\mu} t_{i\alpha ,j \beta}  c^{\dagger}_{i\alpha} c_{j \beta} \\
+ & i\lambda \sum_{i,\mu \nu} (\mathbf{c}^{\dagger}_{i \mu} \times \mathbf{c}^{\dagger}_{i \nu})\cdot \mathbf{s}_{\mu\nu},
\end{split}
\end{equation}
where $c_{i \alpha \mu}^{\dagger}$ are electron creation operators on the $\alpha\left(=p_{x}, p_{y}, p_{z}\right)$ orbital with spin $\mu(=\uparrow, \downarrow)$ at the \textit{i}th site. $\epsilon_{\alpha}$
is the on-site energy of the $\alpha$ orbital. $t_{i \alpha, j \beta}=t_{\alpha \beta}\left(\mathbf{r}_{i j}\right)$ is the Slater-Koster hopping integral which depends on orbital types and the intersite vector $\mathbf{r}_{i j}$ from sites $i$ to $j$ \cite{SK,Harrison} . $\lambda$ is the SOC strength, $\mathbf{c}_{i \mu}^{\dagger}=(c_{i p_{x}}^{\dagger}, c_{i p_{y}}^{\dagger}, c_{i p_{z}}^{\dagger})_{\mu},$
and $\mathbf{s}=(\sigma_{x}, \sigma_{y}, \sigma_{z})$ are the Pauli matrices.

As shown in Fig.~\ref{TCI}(b), numerically calculated energy spectrum with a PBC shows an energy gap, while a set of edge states, which are localized on the boundary [see Fig.~\ref{TCI}(c)], emerge in the gap region for the OBC system.
The existence of bulk energy gap and robust midgap edge states suggest a nontrivial electronic topology. Moreover, as shown in the inset Fig.~\ref{TCI}(b), the orbital-resolved PBC spectrum exhibits signatures of a band inversion between $p_z$ and $p_{x,y}$ orbitals. However, the calculated spin Bott index of this system is $B_s=0$, indicating that it is not a QSH state.
Further inspection shows that the edge states of the system are actually protected by the in-plane mirror symmetry. Therefore, this aperiodic system is actually a mirror-protected TCI.

Following a similar idea of the spin Bott index for nonperiodic systems, we propose the topological invariant called the \textit{mirror Bott index} $B_m$ in analogy to the mirror Chern number \cite{mirror_chern} of periodic systems to characterize the aperiodic TCI.
Specifically, we divide the wave functions in the mirror-invariant plane into two separate sets according to their mirror eigenvalues ($\pm i$), and calculate their respective Bott index $B_{\pm i}$. Then, the mirror Bott index is defined as the half difference between them $B_m=(B_{+i}-B_{-i})/2$.
To do so, we construct the projected mirror operator,
\begin{equation}
P_m=P\hat{M}_zP,
\end{equation}
where $P$ is the projector operator defined in Eq.~(\ref{projector}) and $\hat{M}_z=-i\sigma_z\otimes m_z$ with the Pauli matrix $\sigma_z$ and $m_z$ being the mirror matrices in the spin and orbital spaces, respectively.
By solving the eigenvalue equation,
\begin{equation}
P_m|\phi_j^{\pm}\rangle=\pm i |\phi_j^{\pm}\rangle,
\end{equation}
one can construct projector operators for the two mirror subspaces,
\begin{equation}
P_{\pm i}=\sum_j^{N_{occ}/2}|\phi_j^{\pm}\rangle\langle\phi_j^{\pm}|.
\end{equation}
Then, it is straightforward to calculate the Bott index for each subspace as well as the mirror Bott index, following the same algorithm in Eq.~(\ref{xy})-(\ref{Bs}). The mirror Bott index is topologically equivalent to the mirror Chern number, which gives a $\mathbb{Z}$ classification for mirror-protected aperiodic TCIs.

For the quasicrystal in Fig.~\ref{TCI}(b), we found that $B_m = 2$, confirming its nontrivial topological nature. According to the general bulk-edge correspondence, $B_m = 2$ dictates that there exist two pairs of counterpropagating edge states within the energy gaps. These edge states also lead to a quantized two-terminal conductance of $2e^2/h$ per edge, which is verified by the quantum transport simulation.

\subsection{Topological superconductors}
\noindent
In addition to various insulating topological states mentioned above, there are also some works studying topological superconductors in quasicrystals.
\cite{tezuka2012reentrant,degottardi2013majorana,ghadimi2017majorana,Loring,TSC1,TSC2}.
The effect of quasiperiodic potentials on 1D topological superconductors was studied as early as 2012 \cite{tezuka2012reentrant}.
Later, 2D topological superconductors were also explored in quasicrystals \cite{Loring,TSC1,TSC2}.

In Ref.~\cite{Loring}, Fulga \textit{et al.} consider a 2D topological superconductor 
on an AB tiling quasicrystal. Similar to the Bogoliubov-de Gennes (BdG) Hamiltonian of spinless \textit{p}-wave superconductors, the Hamiltonian on the quasicrystal is expressed as
\begin{eqnarray}
H_{\mathrm{QC}} &=& \sum_j \mathbf{c}^{\dagger}_j H_j \mathbf{c}_j + \sum_{\langle j,k \rangle} \mathbf{c}^{\dagger}_j H_{jk} \mathbf{c}_k,\label{H_QC}\\
H_{j}&=&-\mu \sigma_{z},\\
H_{jk} &=& -t \sigma_z - \frac{i}{2} \Delta \sigma_x \cos(\alpha_{jk}) - \frac{i}{2} \Delta \sigma_y \sin(\alpha_{jk}),\label{Hjk}
\end{eqnarray}
where $\mathbf{c}_j^\dag=(c_j^\dag,c_j)$ contains the fermionic creation and annihilation operators at site $j$, $\mu$ is the chemical potential, $t$ is the hopping strength, $\Delta$ is the strength of the \textit{p}-wave paring and $\alpha_{j k}$ is the angle of the bond between site $j$ and site $k,$ measured with respect to the horizontal direction.

\begin{figure}[hbt]
\centering
\includegraphics[width=0.5\columnwidth]{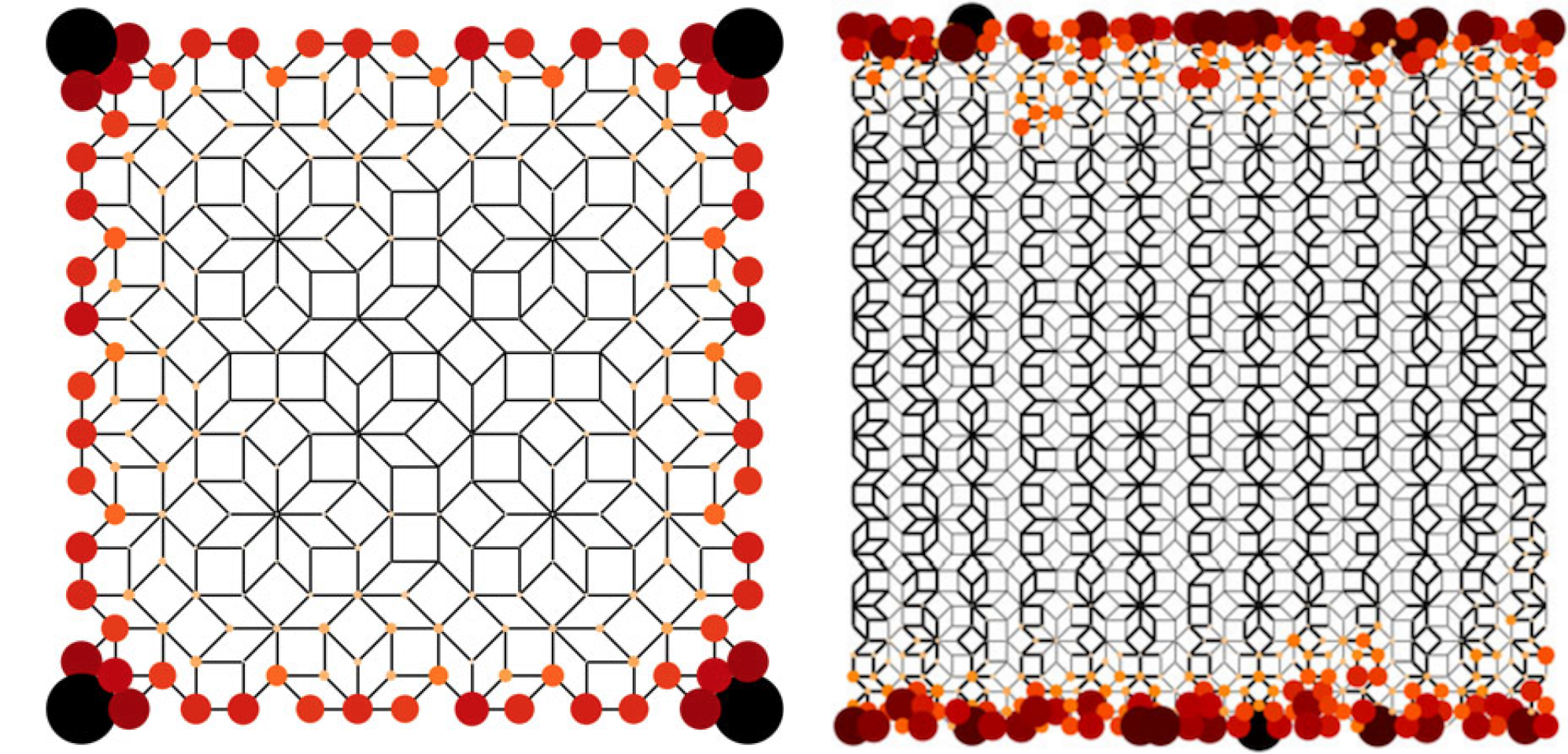}
\caption{(left) Total wave function amplitude of $H_{\mathrm{QC}}$ in the Ammann-Beenker tiling lattice, corresponding to states with energies $|E|<0.2,$ for $t=\Delta=1$ and $\mu=2$. Circles of larger area and darker color correspond to larger amplitudes. (right) Total amplitude of wave functions with energies $|E|<0.1$ for a single disorder realization.Thicker hoppings show the positions of Kitaev chains in the array. Reproduced from Ref.~\cite{Loring}}
\label{TSC}
\end{figure}

The numerical calculation shows gapped bulk and gapless boundary states at the Fermi level for $t = \Delta = 1$ and $\mu = 2$ (see Fig.~\ref{TSC}(a)). The transport properties of the edge states were studied by attaching two infinite translationally invariant leads to the left and right sides of the quasicrystal based on scattering theory \cite{Loring}. For the parameters of Fig.~\ref{TSC}(a), the thermal conductance $G$ is quantized, $G/G_0 = 1$, with $G_{0}=\pi^{2} k_{B}^{2} T_{0} / 6 h$ the quantum of the thermal conductance. This indicates that the quasicrystal is a topological superconductor with Chern number $|C| = 1$.

To confirm the topological superconductor in the quasicrystal, Fulga \textit{et al.} introduce a real-space formulation to determine the topological invariant based on the Clifford pseudospectrum \cite{Loring2}.
Since $H_{\mathrm{QC}}$ obeys particle-hole symmetry
\begin{equation}
\Sigma_{x} H_{\mathrm{QC}} \Sigma_{x}=-H_{\mathrm{QC}}^{*},
\end{equation}
with $\Sigma_{x}=\sigma_{x} \oplus \sigma_{x} \oplus \cdots \oplus \sigma_{x}$, it still belongs to symmetry class $D$ in the Altland-Zirnbauer classification \cite{altland1997nonstandard}. 
Therefore, it allows for a topological classification in terms of the Chern number. 
However, due to the lack of translational symmetry, the topological invariant is obtained as the pseudospectrum $\mathbb{Z}$ index introduced in Ref.~\cite{Loring2}. 
Specifically, one first performs a change of basis, $\tilde{H}_{\mathrm{QC}}=\Omega H_{\mathrm{QC}} \Omega^{\dagger}$ with $\Omega=A \oplus A \oplus \cdots \oplus A,$ and
\begin{equation}
A=\sqrt{\frac{1}{2}}\left(\begin{array}{cc}
1 & 1 \\
-i & i
\end{array}\right),
\end{equation}
so that the Hamiltonian becomes an imaginary one, $\tilde{H}_{\mathrm{QC}}=-\tilde{H}_{\mathrm{QC}}^*$. After that, the \textit{pseudospectrum $\mathbb{Z}$ index} can be obtained as
\begin{equation}
C_{ps} = \frac{1}{2} \mathrm{sig} \begin{pmatrix} X & Y - i \Tilde{H}_{QC} \\ Y + i \Tilde{H}_{QC} & -X \end{pmatrix},
\end{equation}
where $X$ and $Y$ are the position operators associated to the sites of the tiling and $\textit{sig}$ stands for matrix signature, i.e., the number of positive eigenvalues minus the number of negative eigenvalues. The calculated $C_{ps} = -1$, consistent with the quantized values of the thermal conductance. Therefore, the index $C_{ps}$ is a well-defined topological invariant which characterizes 2D topological superconductors belonging to class $D$ in quasicrystals.

In addition, the compatibility of weak topological superconductors with quasicrystals is also explored.
Opposite to strong topological phases which are generally protected by intrinsic symmetries such as time-reversal, particle-hole or chiral symmetries, weak topological phases are usually described in terms of protection by the lattice translational symmetry. It was shown that a weak topological phase can be protected by an averaged translational symmetry and, therefore, can still be robust even when disorder breaks the translational symmetry \cite{PhysRevB.86.045102,PhysRevB.89.155424}. Thus, weak topological superconductors can also be introduced in quasicrystals.


A weak topological superconductor in 2D class $D$ system can be regarded as an array of weakly coupled 1D Kitaev chains which are 1D strong topological superconductors \cite{kitaev2001unpaired}. To convert the 2D AB tiling quasicrystals into an array of Kitaev chains, they selectively reduce the hopping amplitudes in stripe regions of quasicrystals, thus forming an array of coupled quasi-1D strips, as shown in Fig.~\ref{TSC}(b). With this modified Hamiltonian, each quasi-1D strip becomes a nontrivial Kitaev chain hosting Majorana end modes, which forms two Kitaev edges due to weak inter-chain couplings.
As shown in Fig.~\ref{TSC}, a strong topological superconductor exhibits extended edge states along the quasicrystal's boundary, whereas the weak topological superconductor possesses edge states only at the top and bottom boundaries and $C_{ps} = 0$.

To characterize the weak topological superconductor phase in quasicrystals, they introduce a real-space formulation of the weak topological invariant based again on the Clifford pseudospectrum, which is given by
\begin{equation}
Q_{y}=\operatorname{sgn} \operatorname{det}\left(Y+i \tilde{H}_{\mathrm{QC}}\right).
\end{equation}
For the system in Fig.~\ref{TSC}(b), a nontrivial value of $Q_y = -1$ is found, which verified the existence of the weak topological phase in quasicrystals.

\section{Quasicrystalline symmetry-protected topological states}
\noindent
In addition to the realization of quasicrystalline counterparts of topological states already existing in crystals, it is more interesting to explore novel topological states that can only appear in quasicrystals. Without classical crystallographic restriction, quasicrystals may exhibit rotational symmetries that are forbidden in conventional crystals. Here we briefly introduce the recently proposed higher-order topological phase protected by unique quasicrystalline rotational symmetries.

Comparing to the well-studied topological (crystalline) insulators, which have a gapped $n$-dimensional bulk and topologically protected $(n-1)$-dimensional gapless boundary states, a HOTI in $n$-dimension has a gapped $n$-dimensional bulk and $(n-1)$-dimensional boundary, but the gapless states emerge at lower dimensions, e.g., the 0D corner of a 2D insulator. \cite{schindler2018higher, langbehn2017reflection}. As an extension of TCIs, the recently proposed HOTIs are also related to lattice symmetry and have been extensively investigated in crystals.
Recently, the concept of HOTIs were extended to quasicrystals, which results in quasicrystalline symmetry-protected higher-order topological phases \cite{HOTI1,Chen, HOTI2, HOTI3}.



\begin{figure}[hbt]
\centering
\includegraphics[width=0.5\columnwidth]{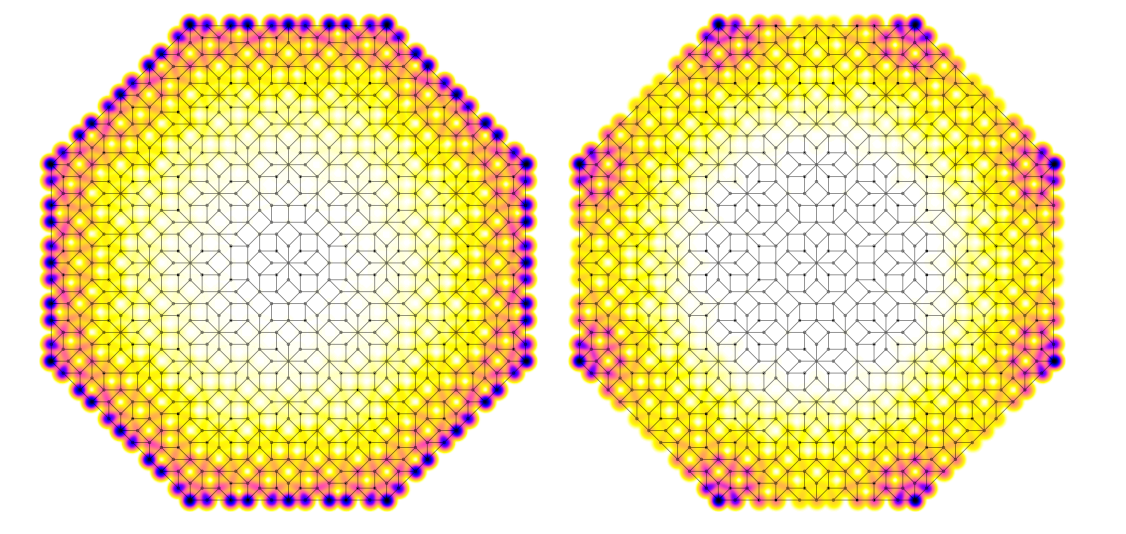}
\caption{The real-space distributions of the wave function amplitudes in the eight lowest energy states of the model defined in Eq.~(\ref{Hhoti}) and (\ref{V8}). (left) $V = 0$, the system hosts counterpropagating Majorana modes on each edge, protected by mirror symmetry. (right) $V = 1$, edge states are gapped out, leading to a HOTI phase. A single Majorana zero mode is localized in each corner of quasicrystals.  Reproduced from Ref.~\cite{HOTI1}}
\label{HOTI}
\end{figure}

For example, Varjas \textit{et al.}~\cite{HOTI1} constructed a 2D higher-order topological superconductor protected by a quasicrystalline eight-fold rotational symmetry, where localized Majorana zero modes bound to the eight corners of the octagonal AB tiling.
Specifically, they start from a tight-binding model describing two oppositely spin-polarized $p \pm ip$ topological superconductors in class $D$ \cite{PhysRevB.61.10267}. 
Similar to Eq.~(\ref{H_QC})-(\ref{Hjk}), the real-space BdG Hamiltonian on the AB tiling quasicrystal is given by
\begin{eqnarray}
\mathcal{H}&=&\sum_{j} \boldsymbol{\Psi}_{j}^{\dagger} \mathcal{H}_{j} \boldsymbol{\Psi}_{j}+\sum_{\langle j, k\rangle} \boldsymbol{\Psi}_{j}^{\dagger} \mathcal{H}_{j k} \boldsymbol{\Psi}_{k},\label{Hhoti}\\
\mathcal{H}_{j}&=&\mu \sigma_{z} \tau_{z},\\
\mathcal{H}_{j k}&=&\frac{t}{2} \sigma_{z} \tau_{z}+\frac{\Delta}{2 i}\left[\cos \left(\alpha_{j k}\right) \sigma_{z} \tau_{x}+\sin \left(\alpha_{j k}\right) \sigma_{z} \tau_{y}\right],
\end{eqnarray}
where $\boldsymbol{\Psi}_{j}^{\dagger}=\left(\psi_{j, \uparrow}^{\dagger}, \psi_{j, \uparrow}, \psi_{j, \downarrow}^{\dagger}, \psi_{j, \downarrow}\right)$, $\psi_{j, \sigma}^{\dagger}$ is the fermionic creation operator at site $j$ with spin $\sigma$, $\langle\cdots\rangle$ denotes sites connected by a bond (see Fig.~\ref{HOTI}). $\mu$ is the chemical potential, and Pauli matrices $\tau$ and $\sigma$ acts on the electron-hole and spin degrees of freedom, respectively. $t$ is the hopping amplitude, $\Delta$ is the \textit{p}-wave pairing strength and $\alpha_{jk}$ is the angle formed by the bond with respect to the horizontal direction.

The system obeys particle-hole symmetry $\{\mathcal{H},\mathcal{P}\}=0$, in-plane mirror symmetry $[\mathcal{H},M]=0$, and particularly a global eight-fold rotational symmetry $[\mathcal{H},C_8]=0$. These symmetry operators are written as: $\mathcal{P}=\tau_x\sigma_0 \mathcal{K},
M=\tau_0\sigma_z,$ and
\begin{equation}
C_8=\exp(-i\frac{\pi}{8}\sigma_0\tau_z)\mathcal{R},
\end{equation}
where $\mathcal{K}$ denotes complex conjugation and $\mathcal{R}$ is an orthogonal matrix permuting the sites of the tiling to rotate the whole system by an angle of $\pi/4$. As the Hamiltonian (\ref{Hhoti}) describes a bilayer system of two 2D class-$D$ topological superconductors with opposite Chern number, a pair of counterpropagating Majorana edge modes, which are prevented from gapping out by $M$, exist on the boundary of the quasicrystal, as shown in Fig.~\ref{HOTI}(a). To obtain a higher-order topological superconductor, they introduce a perturbation that breaks both $M$ and $C_8$, but preserves their product $C8M$, which is given by
\begin{equation}\label{V8}
\mathcal{V}=\sum_{\langle j, k\rangle} \boldsymbol{\Psi}_{j}^{\dagger} \mathcal{V}_{j k} \boldsymbol{\Psi}_{k}, \quad \mathcal{V}_{j k}=\frac{V}{2} \sigma_{y} \tau_{0} \cos \left(4 \alpha_{j k}\right).
\end{equation}
Because $\mathcal{V}$ anticommutes with both $M$ and $C_8$, the edge spectrum is gapped out and the gap changes sign alternatively across
the eight edges of the AB tiling quasicrystal.  As a result, eight Majorana zero modes are formed as domain-wall states at corners of the octagonal sample \cite{JR}, as shown in Fig.~\ref{HOTI}(b).
As an important manifestation of the nontrivial bulk topology of the higher-order topological superconductor, these Majorana zero modes bound to the corners are robust against any perturbations preserving $\mathcal{P}$ and $C_8 M$ symmetries. Because the eightfold rotational symmetry is forbidden in periodic crystals by the crystallographic restriction theorem, the resulting higher-order topological phase has no crystalline counterpart.


It was previously shown that the existence of Majorana bound states on the corners on the 2D topological crystalline superconductors indicates the nontrivial topology of the bulk and a $\mathbb{Z}_2$ index defined at the rotational symmetric points can be constructed to characterize the phase \cite{PhysRevLett.111.047006}. However, since there is no translational symmetry in quasicrystalline lattices, Varjas \textit{et al.}~introduce a momentum-dependent effective Hamiltonian $H_{\mathrm{eff}}=G_{\mathrm{eff}}^{-1}$, with the effective Green's function in the plane-wave bases
\begin{equation}
G_{\mathrm{eff}}(k)_{n, m}=\langle k, n|G| k, m\rangle,
\end{equation}
where $|k, n\rangle$ is the normalized plane-wave state with nonzero amplitude in orbital $n$, and $G = \lim _{\eta \rightarrow 0}(H+i \eta)^{-1}$ is the zero-energy Green's function of the full Hamiltonian. Since the gap of $H_{\mathrm{eff}}$ closes only when the gap of the full Hamiltonain $H$ closes, the topological invariant defined in terms of $H_{\mathrm{eff}}$ can also characterize the topology of the full Hamiltonian.

Since the Majorana bound states are protected by $C_8 M$ and $\mathcal{P}$, the Pfaffian can be computed for each $C_8 M$ invariant subspace related by $\mathcal{P}$. Specifically, $C_8 M$ have eight eigenvalues $\omega_{n}=\exp [i(\pi / 8) n]$ with $n = [\pm1,\pm3,\pm5,\pm7]$, and eigenstates $|n\rangle$ and $|\text{-}n\rangle$ related by $\mathcal{P}$. By restricting $H_{\mathrm{eff}}(k=0)$ to $C_8M$ eigensubspaces of $\omega_{\pm n}$, the signs of the four Pfaffians $\nu_{n, k}=\pm 1,$ for $n \in[1,3,5,7]$ in each subspace represent a $\mathbb{Z}_{2}^{4}$ classification.

However, this is not a stable topological invariant as it also distinguishes different atomic insulators with on-site Hamiltonians of opposite signs and vanishing hoppings. To solve this issue, they invoke the cut-and-project method to build the AB tiling from a 4D ancestor lattice \cite{baake2013aperiodic}, and used the plane-wave states in the 4D lattice as an overcomplete basis for the 2D quasicrystals. Then they consider the effective Hamiltonian at $C_8$ invariant momenta in the 4D lattice, which are just two momenta $\Gamma=(0,0,0,0) \equiv 0$ and $\Pi=(\pi, \pi, \pi, \pi)$. The $\mathbb{Z}_2$ invariant for each subspace $|n\rangle$ and $|\text{-}n\rangle$ is then just $\nu_{n}=\nu_{n, 0} / \nu_{n, \Pi}$. Because this model has $\nu_{1}=\nu_{7}$ and $\nu_{3}=\nu_{5}$, the topological invariant is further simplified as $\nu = \nu_1 \nu_3$, whose nontrivial value characterizes the presence of the corner Majorana zero modes.

Although higher-order topological phases were extended to octagonal quasicrystals \cite{HOTI1, Chen} (as elucidated above) as well as dodecagonal quasicrystals, \cite{HOTI2,HOTI3} the argument for the existence of eight corner Majorana zero modes relies on an alternating sign of the mass terms at the boundary. It fails in quasicrystals with odd-rotational symmetries, e.g., $C_5$. Therefore, it remains unknown whether higher-order topology can exist in quasicrystals with $C_5$, which is the characteristic symmetry of the first experimentally discovered quasicrystal. \cite{PhysRevLett.53.1951}

\section{Summary and outlook}
\noindent In summary, we give a general overview of the recent progress of topological states in quasicrystals. We first introduce fundamental definition of quasicrystals, basic theoretical approaches to modeling quasicrystals, and experimental platforms to realize quasicrystalline structures. Then, we systematically discuss the topological states in 1D and 2D quasicrystals. In 1D quasicrystals, a phasonic degree of freedom, which serves as a synthetic dimension, plays a crucial role in mapping the 1D quasicrystals to 2D IQHEs. Hence, the 1D quasicrystals are associated with the same Chern number that characterizes the 2D IQHEs, and inherits its robust boundary state from the 2D ancestors. In 2D, various topological states existing in crystals are generalized to quasicrystals. However, due to the lack of translational symmetry in quasicrystals, original topological invariants defined for crystalline systems do not apply in quaiscrystals, and different real-space expressions of corresponding topological invariants are introduced to characterize these topological states in quasicrystals. Particularly, with the generalization of HOTIs to quasicrystals, unique quasicrystalline symmetry-protected higher-order topological phases were proposed. As the topological protection of these phases explicitly requires rotational symmetry that is incompatible with any periodic crystal structure, thus the resulting phases have no crystalline counterpart.

The research of topological states in quasicrystals is of great interest to fundamental research. 
However, despite some pioneering work, the research field is far from maturity. There are still many challenging issues to be resolved in the future. First, topological states in 3D quasicrystals are rarely studied. New physical mechanism and rich phenomena are expected in higher dimensions. Second, since quasicrystals exhibit crystallographic forbidden symmetries and unique elementary excitations, searching for new topological effects that are directly related to these novel features of quasicrystals is still ongoing. Moreover, opposite to topological states in crystals, where the topological classification based on crystalline symmetries and internal discrete symmetries have been established \cite{kitaev2009periodic,tenfold_way1,tenfold_way2, bradlyn2017topological, po2017symmetry,song2018quantitative}, however, 
a comprehensive topological classification including quasicrystalline symmetries is as yet lacking. Third, with the rapid development of controllable shaping of various metamaterials for optical, acoustic and matter waves, more exotic topological phenomena in artificial quasicrystal structures are awaiting to be experimentally explored. In this context, searching for new topological states in quasicrystals, exploring emerging new physics, and experimental studying new phenomena will be promising research subjects in the future, which may open exciting possibilities in this fascinating research field.

\section*{Acknowledgement}
This work was supported by the National Natural Science Foundation of China (Grant No. 12074006) and the start-up funds from Peking University. 








\end{document}